\documentclass[
reprint,
superscriptaddress,
showpacs,preprintnumbers,
 amsmath,amssymb,
 aps,
 prc,
]{revtex4-1}
\usepackage{subcaption}
\usepackage{graphicx}
\usepackage{dcolumn}
\usepackage{bm}
\usepackage{ upgreek }
\usepackage{units}
\usepackage[utf8]{inputenc}
\usepackage[T1]{fontenc}
\usepackage[colorlinks=true, allcolors=blue]{hyperref}
\usepackage{multirow}
\usepackage{booktabs}
\usepackage{tabularx}

\graphicspath{{figures/}}

\begin{document}

\preprint{APS/123-QED}

\title{Exploring the mass surface near the rare-earth abundance peak \\via precision mass measurements at JYFLTRAP}

\author{M.~Vilen}
\email{markus.k.vilen@student.jyu.fi}
\affiliation{University of Jyv{\"a}skyl{\"a}, P.O. Box 35, FI-40014 University of Jyv{\"a}skyl{\"a}, Finland}
\author{J.M.~Kelly}
\email{jkelly27@alumni.nd.edu}
\affiliation{University of Notre Dame, Notre Dame, Indiana 46556, USA}
\author{A.~Kankainen}
\affiliation{University of Jyv{\"a}skyl{\"a}, P.O. Box 35, FI-40014 University of Jyv{\"a}skyl{\"a}, Finland}
\author{M.~Brodeur}
\affiliation{University of Notre Dame, Notre Dame, Indiana 46556, USA}
\author{A.~Aprahamian}
\affiliation{University of Notre Dame, Notre Dame, Indiana 46556, USA}
\author{L.~Canete}
\affiliation{University of Jyv{\"a}skyl{\"a}, P.O. Box 35, FI-40014 University of Jyv{\"a}skyl{\"a}, Finland}
\author{R.~de~Groote}
\affiliation{University of Jyv{\"a}skyl{\"a}, P.O. Box 35, FI-40014 University of Jyv{\"a}skyl{\"a}, Finland}
\author{A.~de~Roubin}
\affiliation{University of Jyv{\"a}skyl{\"a}, P.O. Box 35, FI-40014 University of Jyv{\"a}skyl{\"a}, Finland}
\author{T.~Eronen}
\affiliation{University of Jyv{\"a}skyl{\"a}, P.O. Box 35, FI-40014 University of Jyv{\"a}skyl{\"a}, Finland}
\author{A.~Jokinen}
\affiliation{University of Jyv{\"a}skyl{\"a}, P.O. Box 35, FI-40014 University of Jyv{\"a}skyl{\"a}, Finland}
\author{I.D.~Moore}
\affiliation{University of Jyv{\"a}skyl{\"a}, P.O. Box 35, FI-40014 University of Jyv{\"a}skyl{\"a}, Finland}
\author{M.R.~Mumpower}
\affiliation{Theory Division, Los Alamos National Lab, Los Alamos, New Mexico 87544, USA}
\author{D.A.~Nesterenko}
\affiliation{University of Jyv{\"a}skyl{\"a}, P.O. Box 35, FI-40014 University of Jyv{\"a}skyl{\"a}, Finland}
\author{J.~O'Brien}
\affiliation{University of Jyv{\"a}skyl{\"a}, P.O. Box 35, FI-40014 University of Jyv{\"a}skyl{\"a}, Finland}
\altaffiliation{University of Liverpool, Liverpool L69 7ZE, United Kingdom}
\author{A.~Pardo Perdomo}
\affiliation{University of Notre Dame, Notre Dame, Indiana 46556, USA}
\author{H.~Penttil{\"a}}
\affiliation{University of Jyv{\"a}skyl{\"a}, P.O. Box 35, FI-40014 University of Jyv{\"a}skyl{\"a}, Finland}
\author{M.~Reponen}
\affiliation{University of Jyv{\"a}skyl{\"a}, P.O. Box 35, FI-40014 University of Jyv{\"a}skyl{\"a}, Finland}
\author{S.~Rinta-Antila}
\affiliation{University of Jyv{\"a}skyl{\"a}, P.O. Box 35, FI-40014 University of Jyv{\"a}skyl{\"a}, Finland}
\author{R.~Surman}
\affiliation{University of Notre Dame, Notre Dame, Indiana 46556, USA}
\date{\today}

\begin{abstract}
The JYFLTRAP double Penning trap at the Ion Guide Isotope Separator On-Line (IGISOL) facility has been used to measure the atomic masses of 13 neutron-rich rare-earth isotopes. Eight of the nuclides, $^{161}$Pm, $^{163}$Sm, $^{164,165}$Eu, $^{167}$Gd, and $^{165,167,168}$Tb, were measured for the first time. The systematics of the mass surface has been studied via one- and two-neutron separation energies as well as neutron pairing-gap and shell-gap energies. The proton-neutron pairing strength has also been investigated. The impact of the new mass values on the astrophysical rapid neutron capture process has been studied. The calculated abundance distribution results in a better agreement with the solar abundance pattern near the top of the rare-earth abundance peak at around $A\approx165$.

\end{abstract}

\maketitle

\section{\label{sec:intro}Introduction}
\begin{figure}[!]
    \includegraphics[trim = 16mm 5mm 0mm 17mm, clip,width=\linewidth]{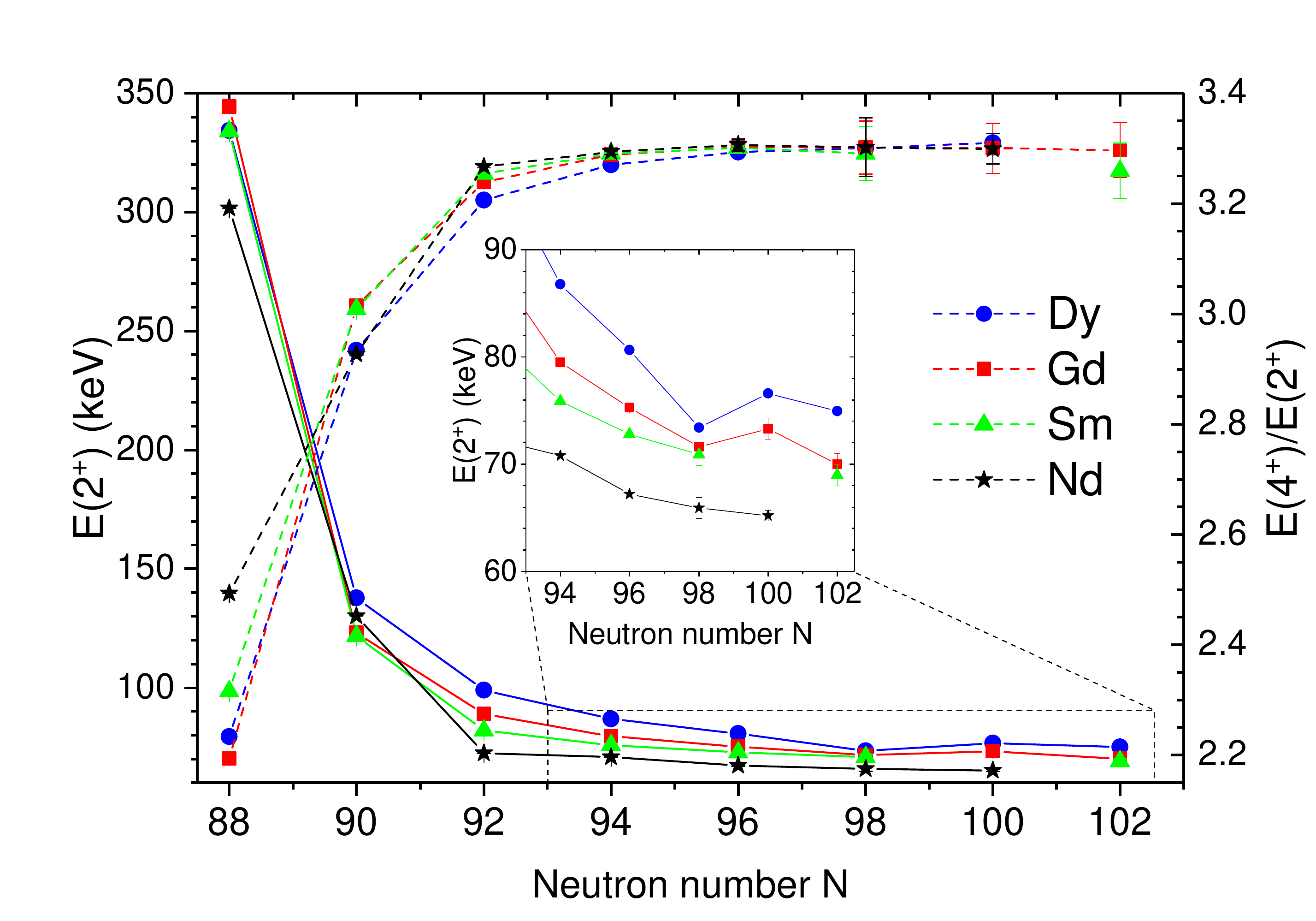}	
    \caption{Experimental excitation energies of the first $2^+$ states (solid lines) together with the ratio of the first $4^+$ and $2^+$ states (dashed lines). The inset shows a peak in $2^+$ energies at $N=100$. The energies have been adopted from ENSDF \cite{ensdf} and \cite{Patel2014}.}
    \label{fig:e2e4}
\end{figure}

The rare-earth region near $A$ = 165 is of interest for both nuclear structure and nuclear astrophysics. With regards to nuclear structure, an onset of strong prolate deformation at $N=88-90$ in these isotopic chains was discovered already in the 1950s \cite{Brix1952,Mottelson1955}. The rapid shape change can also be observed in the excitation energies of the first $2^+$ and $4^+$ states (see Fig.~\ref{fig:e2e4}). The $2^+$ excitation energies decrease strongly after $N=88$, and $E(4^+)/E(2^+)$ ratios reach $\approx 3.3$, compatible with a rigid rotor. A possible subshell closure at $N$ = 100 has been proposed to explain recent experimental data \cite{Jones2005,Ghorui2012,wu2017,Patel2014,Yokoyama2017}, such as the peak in the $2^+$ energies at $N=100$ (see Fig.~\ref{fig:e2e4}). There are also indications of rapid nuclear shape transitions in Nd isotopes \cite{gupta2015}. More recently, an unusual change in nuclear structure at $N$ = 98 near europium has been identified \cite{hartley2018}, and interpreted as a deformed subshell gap. In this work, we investigate whether such structural changes are also observed in nuclear binding energies.

 Neutron-rich rare-earth isotopes play an important role in the astrophysical rapid neutron capture process, the $r$ process \cite{Burbidge1957,Cameron1957,Arnould2007,Horowitz2019}. The $r$ process takes place at least in neutron-star mergers as evidenced by the binary neutron star event GW170817 \cite{Abbott2017,Abbott2017b} in August 2017 and its afterglow known as a kilonova \cite{Tanvir2017,Kasen2017}. During the observational period of a few days, the observed kilonova changed from blue to red. The latter color has been interpreted to be due to lanthanide-rich ejecta with high opacities, i.e. heavier ($A>140$) $r$-process nuclides \cite{Tanvir2017,Kasen2017}. To understand the produced abundances of lanthanides in different astrophysical conditions, masses of the involved nuclei have to be known for reliable calculations as they are one of the key inputs for the $r$ process modelling. 
 
The $r$ process produces rare-earth nuclei during its freezeout stage when material is decaying back to stability \cite{Surman1997,mumpower2012_RE}. Nuclear deformation in the $A$ = 165 region is essential in this process as it is reflected in nuclear binding energies and therefore in the behavior of neutron-separation energies. This will consequently affect the neutron-capture and beta-decay rates and steer the reaction flow toward the midshell, creating what is known as the rare-earth abundance peak at around $A$ = 165 \cite{Surman1997,mumpower2012_RE}. Another mechanism producing rare-earth nuclei and its abundance peak in the $r$ process is fission cycling from heavier nuclei \cite{Goriely2013b}.

\begin{figure}[tpb]
	\begin{center}
    	\centerline{\includegraphics[trim = 25mm 10mm 35mm 20mm, clip,width=\linewidth]{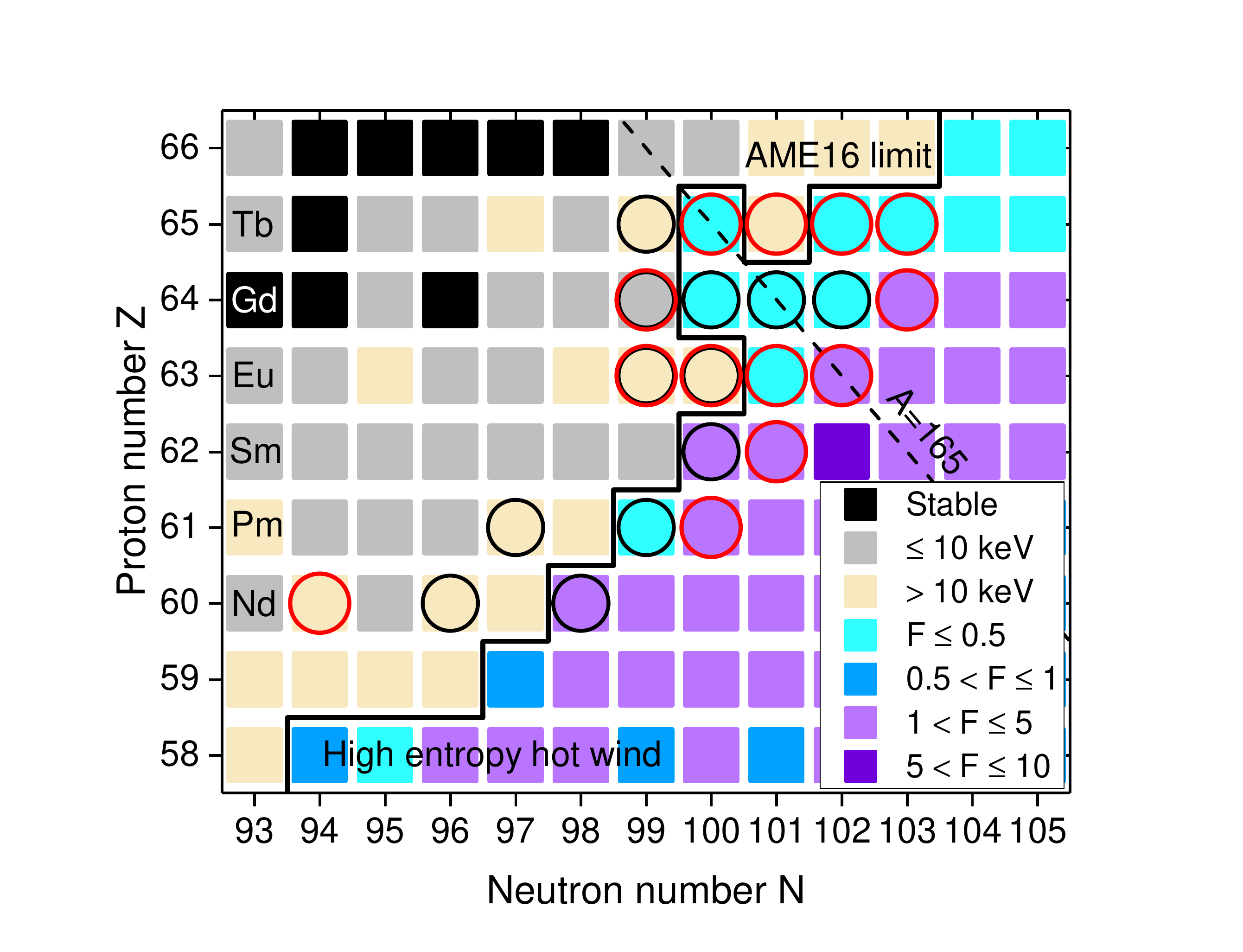}}
    	\caption{The $r$-process impact factors $F$ of masses in the region of interest \cite{mumpower2015prc035807}. Mass measurements from this work are circled in red, while black circles indicate those from an earlier study at JYFLTRAP \cite{vilen2018}. In total, 22 ground-state masses and two isomeric states in this region have been measured at JYFLTRAP, of which 14 go beyond the limit of known nuclei in AME16 \cite{AME16}. For experimental AME16 values, the mass-excess uncertainties have been indicated.}
    \label{fig:chartmasses}
	\end{center}
\end{figure}

The impact of individual nuclear masses on calculated $r$-process abundances can be quantified using a so-called $F$-factor, $F=100\sum_i|X(A)-X_b(A)|$, where $X_b(A)$ is the final isobaric mass fraction in the baseline simulation done with the experimental Atomic Mass Evaluation 2016 (AME16) \cite{AME16} and theoretical FRDM12 \cite{Moller2016} mass values, and $X(A)$ is the final isobaric mass fraction of the simulation when all nuclear inputs have been modified based
on the change in a single mass in an $r$-process simulation \cite{mumpower2015prc035807, mumpower2015prc035807,Mumpower2016}, done using a chosen astrophysical trajectory. Figure~\ref{fig:chartmasses} shows the impact factors in the rare-earth region of interest.

The advent of new and upgraded radioactive ion beam facilities, such as CARIBU \cite{Caribu} and IGISOL-4 \cite{moore2013}, has resulted in a resurgence of mass measurements aimed at characterizing the rare-earth abundance peak \cite{orford2018,hartley2018,vilen2018}. The Canadian Penning Trap (CPT) at CARIBU has measured masses of several Nd, Sm, Pm and Eu isotopes \cite{vanschelt2012,hartley2018,orford2018}. These masses agreed surprisingly well with the predictions of a reverse-engineering mass model \cite{Mumpower2017} that uses the observed shape of the rare-earth abundance peak to predict masses near $A$ = 165. At the new IGISOL-4 facility, the first measurement campaign on the masses of neutron-rich rare-earth isotopes with the JYFLTRAP double Penning trap \cite{Eronen2012} covered 12 relatively high-impact masses, 6 of which were measured for the first time \cite{vilen2018}. The new JYFLTRAP measurements resulted in a smoothening of the calculated $r$-process abundance pattern making it closer to the observed solar pattern. After the successful first campaign at JYFLTRAP \cite{vilen2018}, a second campaign of mass measurements was launched aiming to better understand the formation of the rare-earth abundance peak in the $r$-process as well as the underlying changes in nuclear structure, in particular beyond $N=100$, which has been proposed as a subshell closure by Hartree-Fock calculations \cite{Satpathy2003,Satpathy2004,Ghorui2012}. In this article, we report on the results of the second measurement campaign at JYFLTRAP and study the impact of JYFLTRAP measurements on the $r$ process and nuclear structure in this neutron-rich rare-earth region.
      
\section{\label{sec:exp}Experimental method}

\subsection{Production of neutron-rich rare-earth isotopes at IGISOL \label{sec:igisol}}

The JYFLTRAP double Penning trap mass spectrometer \cite{Eronen2012} is located at the Ion-Guide Isotope Separator On-Line (IGISOL) facility \cite{igisol,moore2013} in the JYFL Accelerator Laboratory of the University of Jyv\"{a}skyl\"{a} in Finland. The neutron-rich rare-earth nuclei of interest were produced through proton-induced fission on uranium at IGISOL, using 25 MeV, 10-15 $\mu$A proton beam from the K-130 cyclotron impinging into a 15 mg/cm$^2$ thick $^{nat}$U target. This target is sufficiently thin to allow the energetic fission fragments to exit out of the target to the target chamber filled with helium at around 300 mbar, and pass through a nickel separation foil to the stopping and extraction volume of the IGISOL fission ion guide \cite{Penttila2012,AlAdili2015}. The thermalized ions are then extracted out of the gas cell by employing a sextupole ion guide \cite{spig} and differential pumping. The extracted ion beam is subsequently accelerated to an energy of $30q$kV, where $q$ is the charge of the ion, and non-isobaric contaminants are separated using a dipole magnet with a mass resolving power ($M/\Delta M$) of about 500. Finally, prior to the injection into JYFLTRAP, the ion beam is decelerated, accumulated, cooled and bunched using a segmented radio-frequency quadrupole ion trap \cite{rfq}.

\subsection{Mass measurements with JYFLTRAP \label{sec:jyfltrap}}
\subsubsection{JYFLTRAP double Penning trap mass spectrometer}
The JYFLTRAP double Penning trap is comprised of two orthogonalized \cite{GABRIELSE19841} cylindrical Penning traps located in the common bore of a 7~T superconducting solenoid \cite{kolhinen2004,Eronen2012}. The first trap, known as the purification trap, is gas-filled and used to remove isobaric contaminants via the sideband cooling technique \cite{Savard1991}. This technique alone can usually provide sufficient cleaning by mass-selectively converting ion motion from magnetron to reduced cyclotron motion. This leads to the centering of the ions in the trap after collisions with the buffer gas. Only the centered ions will be extracted through the 1.5 mm aperture separating the purification and the high-vacuum second trap known as the precision trap. When a sample demands higher resolving power for selecting the ions of interest, then the Ramsey cleaning technique \cite{r-cleaning} is employed following sideband cooling. Here, the ions extracted through the aperture to the second trap undergo an additional cleaning step utilizing a dipolar excitation at the reduced cyclotron frequency ($\nu_+$), which selectively increases the cyclotron radius. A subsequent transfer back to the first trap through the aperture leaves contaminants implanted on the diaphragm.

The ion's cyclotron frequency $\nu_c = qB/(2\pi M)$, where $B$ is the magnetic field strength, $q$ is the charge and $M$ the mass of the ion, is determined in the second Penning trap.  Conventionally, the time-of-flight ion-cyclotron-resonance (TOF-ICR)  \cite{Graff1980,Konig1995} technique has been used to determine $\nu_c$ with either a single quadrupole excitation or the so-called Ramsey excitation \cite{ramsey1,ramsey2}. The latter is comprised of two short excitation pulses separated by a period without excitation. The method can result in a three-fold gain in precision. The TOF-ICR technique has been used exclusively for mass measurements at JYFLTRAP until 2018 when the newer phase-imaging ion-cyclotron-resonance (PI-ICR) technique \cite{Eliseev2014} was successfully commissioned and implemented at JYFLTRAP \cite{Nesterenko2018}. The measurements using the two methods in this work are further described in the following subsections \ref{sec:tof-icr} and \ref{sec:pi-icr}.

\subsubsection{TOF-ICR measurements \label{sec:tof-icr}}
In the TOF-ICR technique \cite{Graff1980,Konig1995}, the ion's initial magnetron motion is converted into cyclotron motion by applying a quadrupole excitation pulse with a fixed duration and amplitude at a frequency around the expected cyclotron frequency. This results in a more-or-less complete conversion of the slow magnetron motion into the fast reduced cyclotron motion depending on the excitation frequency. At the resonance frequency, a maximum conversion is achieved resulting in an increase in the associated radial energy, observed as a much shorter time-of-flight of the ions from the precision trap onto a micro-channel plate (MCP) detector located after the trap.

The choice of employed excitation scheme depends on the half-life, production rates and possible presence of isomeric contamination. The higher-precision two-pulse Ramsey technique was applied when the production rate was sufficient. If the production rate was too low, the conventional one-excitation pulse (referred to as ``quadrupolar'') was used. Since a Ramsey excitation cannot easily resolve isomeric states, a long, 1600 ms quadrupolar excitation was used for two cases, $^{163}$Gd and $^{162}$Eu, for which isomeric states have previously been observed \cite{hayashi2014}. The length of the applied single-pulse quadrupolar excitation varied from $\unit[200]{ms}$ used for $^{154}\mathrm{Nd}$, $^{161}\mathrm{Pm}$, and $^{163}\mathrm{Sm}$, to $\unit[400]{ms}$ applied for $^{167}\mathrm{Gd}$, and $\unit[1600]{ms}$ for $^{162}\mathrm{Eu}$, $^{162m}\mathrm{Eu}$, and $^{163}\mathrm{Gd}$. 

\begin{figure}
	\begin{center}
    	\centerline{
    	\includegraphics[trim = 0mm 0mm 0mm 0mm, clip,width=\linewidth]{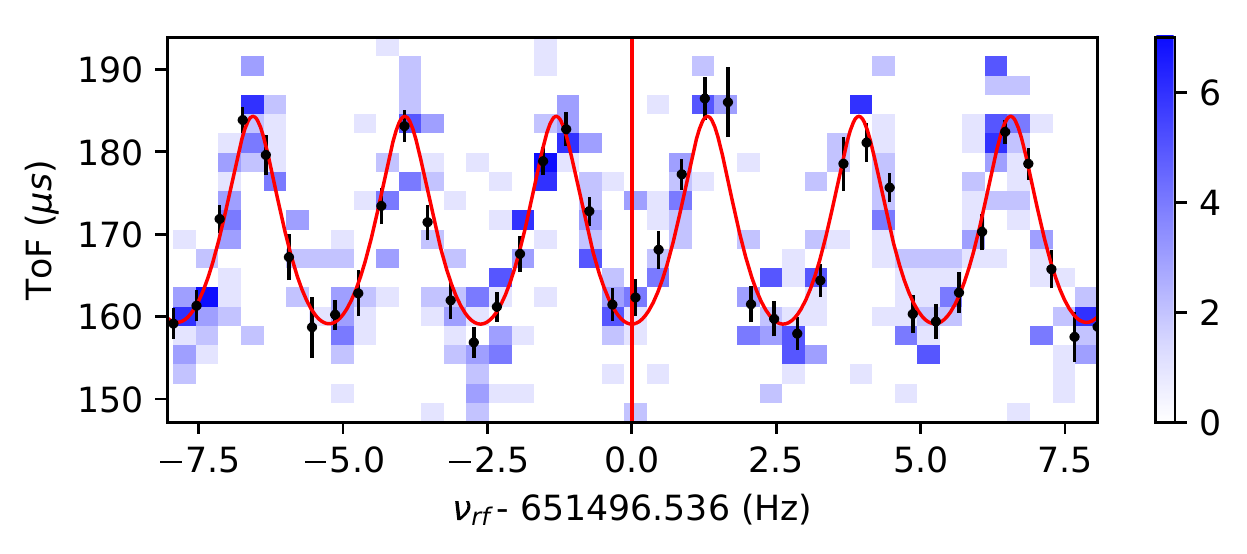}}    	
			\caption{Time-of-flight spectrum for $^{165}$Eu$^+$ using a $\unit[25-350-25]{ms}$ (On-Off-On) Ramsey-type excitation pattern. Background shading indicates the total number of ions, where darker shading indicates more ions. The red line is a fit of the lineshape to the data points (in black).}
    \label{165Eu_TOF-ICR}
	\end{center}
\end{figure}

The remaining TOF-ICR measurements utilized the Ramsey method of time-separated oscillatory fields with two excitation pulses, each with a rectangular envelope. A TOF-ICR resonance for $^{165}$Eu is presented in Fig.~\ref{165Eu_TOF-ICR}. $^{163,165}\mathrm{Eu}$ used an On-Off-On pattern of $\unit[25-350-25]{ms}$ and $^{164}\mathrm{Eu}$ and $^{165-168}\mathrm{Tb}$ used a pattern of $\unit[25-~750-~25]{ms}$.

\subsubsection{PI-ICR measurements\label{sec:pi-icr}}

In the PI-ICR method, determination of the cyclotron frequency relies on detecting the projections of the ion's radial motions in a trap onto a position-sensitive MCP detector with a delay-line anode. Measurements were performed using the second measurement scheme presented in \cite{Eliseev2014}. In this scheme the cyclotron frequency is determined as a sum of magnetron and modified cyclotron frequencies, $\nu_c = \nu_- + \nu_+$, in such a way that the ion's position is recorded for only one phase (referred to as ``phase spot'') for each type of motion, in addition to measuring the center position. The cyclotron frequency was then determined using
\begin{equation}
\nu_c = \frac{\alpha + 2\pi (n_+ + n_-)}{2\pi t}\text{,}
\label{eq:PI-ICR}
\end{equation}
where $\alpha$ is the angle between phase spots of modified cyclotron and magnetron motions, $n_+$ and $n_-$ are the number of revolutions the ion completed in the precision trap during the respective motion and $t$ is the phase accumulation time ions spent in the precision trap. In this work, the PI-ICR technique was used only for $^{162}$Eu$^+$ ions. A phase accumulation time of $\unit[600]{ms}$ was used for all measurements.

\begin{figure}
	\begin{center}
    	\centerline{
    	\includegraphics[trim = 0mm 0mm 0mm 0mm, clip,width=\linewidth]{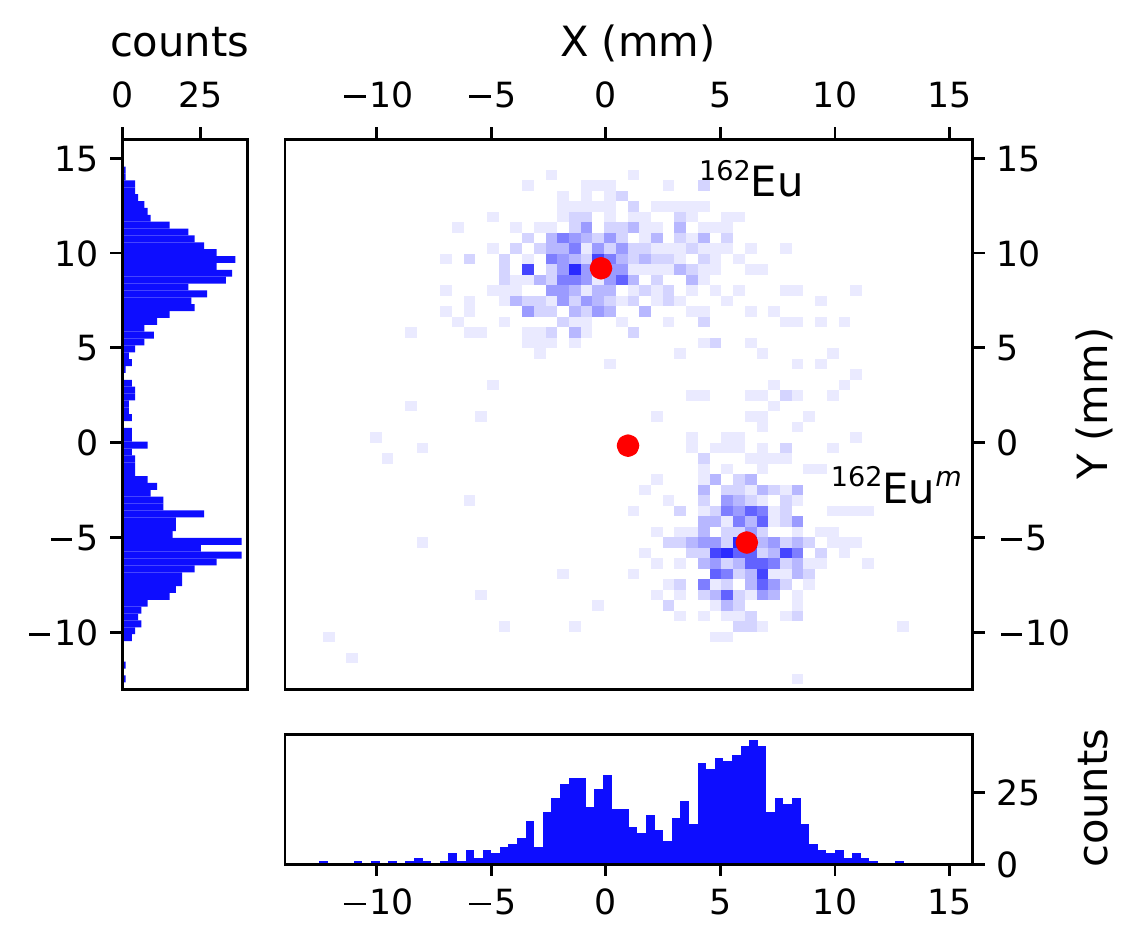}}    	
			\caption{Projection of cyclotron motion of $^{162}$Eu$^+$ ions on the position-sensitive detector detector using the PI-ICR technique. Two detected ion spots correspond to the ground and isomeric state of $^{162}$Eu. In the middle figure background shading indicates the total number of ions, where darker shading indicates more ions. Center point of the two states and the center of the precision trap are marked as red dots. Figures on the left and bottom present projections of the middle figure onto the $Y$ and $X$ axes, respectively.}
    \label{162Eu_PI-ICR}
	\end{center}
\end{figure}

The measurement process was identical to TOF-ICR measurements until the ions were transferred into the precision trap. There, the residual coherent components of axial and radial eigenmotions with frequencies $\nu_z$ and $\nu_-$ were cooled using dipolar excitation pulses with suitable amplitude, phase and frequency. This was followed by an increase of reduced cyclotron eigenmotion amplitude via a dipolar excitation with the $\nu_+$ frequency. Using the two timing patterns, as described in \cite{Eliseev2014}, the ions were given time to accumulate a phase angle, either with frequency $\nu_+$ or $\nu_-$ before being extracted from the precision trap with a non-zero magnetron motion amplitude. This process resulted in at least one spot on the position-sensitive MCP detector for each timing pattern. Additionally, the center of the precision trap was projected onto the detector for determination of the angle between measured spots resulting from the used two timing patterns. In the case of $^{162}$Eu, both the ground state and the isomer were injected into the precision trap simultaneously. The two states were distinguished by allowing a sufficiently long phase accumulation time to pass so that the two states produced separate $\nu_+$ spots on the detector. The two states are presented in Fig.~\ref{162Eu_PI-ICR}.

Spot positions on the MCP were averaged over any residual radial eigenmotion after the initial magnetron cooling in the precision trap and following the conversion from reduced cyclotron motion into magnetron motion before extraction of ions onto the detector. This was achieved via scanning the timing pattern over the period of the relevant eigenmotion such that an equal amount of data was gathered with all steps of the scans. Data resulting from this two-dimensional scan was then used in further analysis. This process is highly beneficial since it enables the measurement and data analysis to be performed without a need to do any scanning or fitting, with the exception of unavoidable extrapolation during count rate analysis. The measurement process merely averages results in accounting for residual eigenmotions.

\subsection{Production of stable reference ions \label{ref:offline}}

Both the TOF-ICR and PI-ICR methods rely on measuring well-known reference ions before and after the ion of interest in order to calibrate the magnetic field strength. This is done by interpolating the measured reference-ion cyclotron frequencies to the time of the ion-of-interest measurement.

In this work, singly-charged $^{136}\mathrm{Xe}$ and $^{133}\mathrm{Cs}$ ions were used as references. $^{136}\mathrm{Xe}$ was used as a reference for most of the studied nuclides, since its mass is well-known and it is readily available as a fission fragment from IGISOL. However, obtaining the reference as a by-product of fission entails a risk of misidentifying it for another species or molecule with nearly equal mass. In order to eliminate this risk, an offline ion source station \cite{vilen2019}, completely separate from the IGISOL target chamber, was later added to the beam line. The stable $^{133}\mathrm{Cs}^+$ reference ions were produced at this offline station.  

The new off-line ion source station \cite{vilen2019} consists of multiple ion sources and a beamline that connects to the existing IGISOL beamline just before the IGISOL dipole magnet. A deflector allows for rapid switching between the off-line ion beams and the radioactive ion beams from the IGISOL target chamber. In this work, a thermal emission alkali metal ion source consisting of $^{39}\mathrm{K}$, $^{85}\mathrm{Rb}$, $^{87}\mathrm{Rb}$, and $^{133}\mathrm{Cs}$ was used. Thanks to the IGISOL dipole magnet mass separator, pure beams of $^{133}\mathrm{Cs}^+$ ions were guaranteed.

\section{\label{sec:analysis}Analysis}

\subsection{\label{sec:RandM}Determination of the atomic masses}

Atomic masses of the measured isotopes were determined using 
\begin{equation}
m = \overline r \cdot (m_{ref}-m_{e}) + m_{e}\text{,}
\end{equation}
where $\overline r$ is the weighted mean of the frequency ratios between the reference and the ion of interest (\( r = \nu_{c,ref}/\nu_{c} \)), $m_{ref}$ is the atomic mass of the reference ion, and $m_e$ is the mass of the electron. The electron binding energies, all in the eV range, are several orders of magnitude smaller than the statistical uncertainty of the measurements, and can therefore be neglected.

\begin{figure}[tpb]
\begin{center}
\centerline{\includegraphics[trim = 0mm 0mm 0mm 0mm, clip,width=\linewidth]{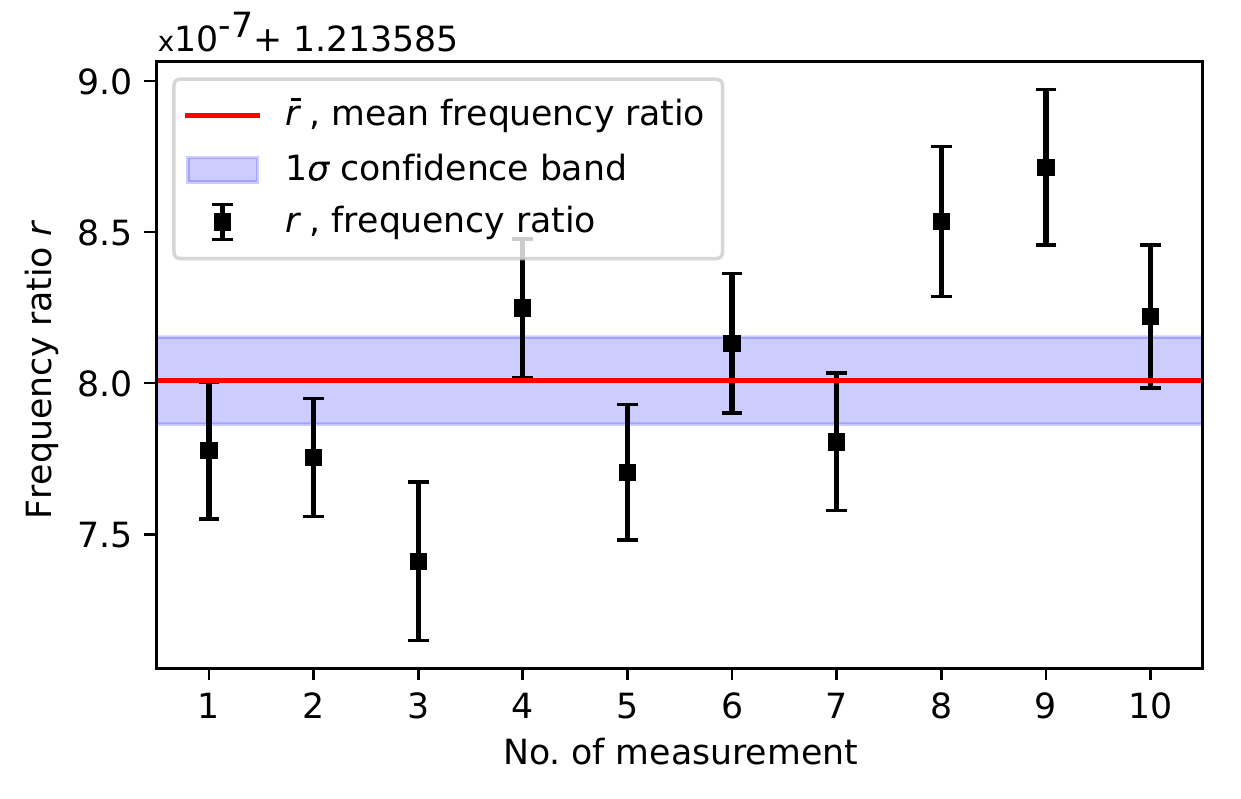}}    	
\caption{(Color online) Measured frequency ratios $r=\nu_c(^{136}$Xe$^+)/\nu_c(^{165}$Tb$^+$) for $^{165}$Tb (black data points) together with the weighted mean (solid red line) and its error band (dashed blue lines). The Birge ratio for $^{165}$Tb measurements was 1.05.}
\label{fig:r_165Tb}
\end{center}
\end{figure}

Multiple frequency ratios $r$ were measured. The weighted mean of the frequency ratio $\overline r$ was determined together with its internal and external uncertainties. An example is shown for $^{165}$Tb$^+$ ions in Fig.~\ref{fig:r_165Tb}. Following the procedure of the Particle Data Group \cite{PDG}, the statistical uncertainty of any $\overline r$ has been inflated by the ratio of the external-to-internal uncertainty (so-called Birge ratio) if its value is greater than one. This practice ensures that possible systematic effects bringing the Birge ratio above the statistically-expected value of one are taken into account. Systematic uncertainties were taken into account and added quadratically to the measured frequency ratios as described in the following section, Sect.~\ref{sec:systematic}.

\subsection{\label{sec:systematic}Systematic uncertainties}

\subsubsection{Uncertainties due to geometry and $B$ field fluctuations\label{sec:Bfield}}
Multiple systematic effects can be present in Penning-trap measurements. These can be due to e.g. misalignment of the trap with respect to the magnetic field, inhomogeneous magnetic field, harmonic and non-harmonic distortion of the trapping potential, and other unavoidable geometric irregularities \cite{brown1986}. The above effects can cause a so-called mass-dependent systematic error on the frequency ratio that ultimately leads to cyclotron frequency determinations away from the expected $\nu_c \propto 1/m$ relationship. At JYFLTRAP, the mass-dependent error has been determined to be equal to $(2.2 \times 10^{-10}/u) \times \Delta m$ \cite{vilen2018}, where $\Delta m$ is the mass difference between the ion of interest and the reference ion. This leads to uncertainties of less than 1 keV for $\Delta m \approx 30 ~u$ in the studied mass region. Another sub-keV systematic error is due to relativistic effects \cite{Brodeur2009}, which are minimal for heavier nuclei such as the ones measured in this work. The mass-dependent uncertainties were added in quadrature to the final frequency ratio uncertainty. 

An additional systematic effect lies in non-linear temporal fluctuations of the magnetic field on top of its slow linear drift over time. This means that even the very shortest linear interpolations between successive reference measurements to determine the magnetic field strength at the time of the interposed ion-of-interest measurement are subject to an error. For JYFLTRAP, these uncertainties have been measured to be $8.18(19) \times 10^{-12}$ min$^{-1} \times \Delta t$ \cite{canete2016}, where $\Delta t$ is the time between consecutive reference measurements. The effect of these fluctuations on the measured frequencies is, however, negligible over the typical time spans between successive reference measurements for this work. 

\subsubsection{\label{sec:ion-ion}Uncertainties due to ion-ion interactions}

The ions are injected into the precision Penning trap as bunches containing typically a few ions. One of the main effects that has to be considered in the data analysis is the ion-ion interactions when multiple ions are simultaneously trapped \cite{Bollen1992}. The effect is especially significant if more than one ion species are trapped but it can also be present when trapping multiple ions of a single species. Most of the measured isotopes had insufficient ion count rates to correct the result for the effect of having multiple ions in the trap. All data measured with the TOF-ICR technique were limited to 1-3 ions per bunch in the analysis phase, and processed without taking the ion-ion interactions into account. An additional systematic uncertainty was added to all obtained frequency ratios to account for any possible ion-ion interactions.

The magnitude of the correction due to ion-ion interactions was determined by analyzing the $^{165}\mathrm{Tb}$ dataset both with and without taking the ion-ion interactions into account. The ion-ion interaction effects were studied by dividing the data into classes of one, two or three detected ions per bunch. The three classes were analyzed separately and the number of ions in each class was corrected with the detection efficiency. The resulting $\nu_c$ frequencies were then extrapolated to the ideal case with only one ion in the trap. A weighted mean was calculated and the larger of the internal and external uncertainties was chosen as the error of the mean. The comparison between the results obtained with and without the count-rate class analysis for the data limited to 1-3 ions per bunch, yielded a small shift of $\delta r/r = (2.2 \pm	3.0)\cdot 10^{-8}$ between the results. Since the shift is compatible with zero within $1\sigma$, we also report the values without this additional systematic contribution in Table~\ref{results}. The additional systematic uncertainty due to the ion-ion interactions was added quadratically to all TOF-ICR results, and is included in the total uncertainty $\delta r_{tot}$ given in Table~\ref{results}. In the case of PI-ICR measurements, the ion count rate was sufficient to take the ion-ion interactions into account. There, the presence of different ion species (ground and isomeric states) simultaneously in the trap turned out to produce a clear shift in the results. 

A detailed characterization of the efficiency of the position sensitive MCP detector and related data acquisition system was conducted in order to minimize any residual systematic uncertainties in the count-rate class analysis. The MCP detector was found to suffer from gradually decreasing efficiency with higher number of ions per bunch. An assumption of a linear relationship between the efficiency-corrected number of ions per bunch and the extrapolated quantity, cyclotron frequency $\nu_c$, or spot position, was assumed as in Ref.~\cite{Kellerbauer2003}. The use of higher order polynomials or other functions with a larger number of fitted parameters was considered, but had to be rejected due to the lack of data with large numbers of ions per bunch. More details on the systematic uncertainties and determination of the detection efficiency can be found on Ref.~\cite{JYFLTRAP_systematics}.

\section{\label{sec:results}Results}

\subsection{Mass-excess values}
 The results of this work are summarized in Table~\ref{tab:results} and Fig.~\ref{fig:all-measurements}. Altogether 13 different nuclides were measured, of which eight were measured for the first time (see Fig.~\ref{fig:chartmasses}). In addition, isomeric states in $^{162}$Eu and  $^{163}$Gd were studied. In the following, the eight nuclides measured for the first time are discussed initially. This is followed by a nuclide-by-nuclide discussion for the other measured nuclides.

\begin{table*}
\caption{\label{tab:results} Frequency ratios ($r=\nu_{c,ref}/\nu_{c}$) based on $N_{meas}$ measurements together with the mass-excess values ($ME$) determined in this work. For the JYFLTRAP values, the uncertainties for $r$ and $ME$ are given both without and with the added systematic uncertainty due to ion-ion interactions, see Sect.~\ref{sec:ion-ion}. Comparison to AME16 \cite{AME16} is given, and a \# sign indicates extrapolated values therein. The isomeric-state mass values were adopted from NUBASE16 \cite{nubase16}. All measurements were done with singly-charged ions. The masses for the reference ions $^{136}\mathrm{Xe}^+$ and $^{133}\mathrm{Cs}^+$ were adopted from AME16 \cite{AME16}. For comparison, the recent CPT measurements are also tabulated.
}
\begin{ruledtabular}
\begin{tabular}{llllllll}
\multirow{2}{*}{Isotope} & \multirow{2}{*}{Reference}&  \multirow{2}{*}{$\mathrm{N}_{meas}$}  & \multirow{2}{*}{$r$} & \multicolumn{4}{c}{$ME \mathrm{(keV)}$}\\
\cmidrule(lr){5-8}
&&&& $JYFL$ & $AME16$ & $\Delta ME$\footnotemark[1] & $CPT$ \\ 
\midrule
$^{154}$Nd 					    & $^{133}$Cs & 3 						& 	1.158 189 215(201)(203)	&	-65601.2(24.9)(25.1)		&	-65825(53)			& 	224(59)			& 	-65579.6(1.0)\cite{orford2018}		
\\
$^{161}$Pm 						& $^{136}$Xe & 5 						& 	1.184 236 679(468)(468)	& -50107.6(59.2)(59.3)		& -50235(298)\# 	& 127(304)\# 	& N/A 	
\\
$^{163}$Sm 						& $^{136}$Xe & 4 						& 	1.198 949 148(285)(286)	& -50552.3(36.0)(36.2)		& -50720(298)\# 	& 168(301)\#	& -50599.6(7.3)\cite{orford2018}	
\\
$^{162}$Eu\footnotemark[2] 						& N/A\footnotemark[2] & 5							&	N/A\footnotemark[2]					&	-58720.4(3.1)\footnotemark[2]		&	-58703(35)				&  -14(36)				&	-58723.9(1.5)\cite{hartley2018}	\\
$^{162}$Eu$^m$\footnotemark[2] 					& $^{133}$Cs & 5								&	1.218 439 459(13)\footnotemark[2]					&	-58565.7(7.6)\footnotemark[2]		& -58540(40)		&  		-20(50)		&	-58563.7(1.9)\cite{hartley2018}	\\
$^{163}$Eu						& $^{133}$Cs & 5						& 		1.225 979 710(14)(30)				&	-56575.7(1.8)(3.8)		&	-56485(66)				&  -91(66)				&	N/A			
\\
$^{164}$Eu						& $^{136}$Xe & 4						&  1.206 285 979(12)(29)		& -53231.1(1.6)(3.7)		&-53381(114)\# 		& 150(115)\# 	& N/A 		
\\
$^{165}$Eu 						& $^{136}$Xe & 3						& 	1.213 663 750(39)(48) 	&	-50726.9(5.0)(6.0) 	& -50724(138)\# 	& -3(139)\#  	& N/A 	
\\
$^{163}$Gd						& $^{136}$Xe & 5						& 	1.198 863 600(76)(81) 	&	-61382.4(9.6)(10.2)		& -61314(8) 		& -68(14) 		& -61316.0(15.0)\cite{vanschelt2012}		
\\
$^{163}$Gd$^m$ 				& $^{136}$Xe & 5 						& 	1.198 864 872(102)(106) 	& -61221.3(13.0)(13.4)		& -61176(8)	&  -45(16) 	& N/A	
\\
$^{167}$Gd 						& $^{136}$Xe & 5 						& 	1.228 379 286(93)(97) 	& -50783.4(11.8)(12.3)		& -50813(298)\# 	&  30(299)\# 	& N/A	
\\
$^{165}$Tb 						& $^{136}$Xe & 10						& 	1.213 585 800(15)(31)	&	-60595.1(2.0)(3.9)		& -60566(102)\# 	& -29(103)\# 	& N/A	
\\
$^{166}$Tb 						& $^{136}$Xe & 6						& 	1.220 965 810(11)(30) 	&	-57807.6(1.6)(3.7)		& -57885(70) 		& 77(71) 		& N/A	
\\
$^{167}$Tb 						& $^{136}$Xe & 5 						& 	1.228 338 998(13)(30)	& -55883.7(1.7)(3.8)		& -55927(196)\# 	& 43(197)\# 	& N/A	
\\
$^{168}$Tb 						& $^{136}$Xe & 5						& 	1.235 721 496(17)(33)	& -52781.2(2.3)(4.1)			& -52723(298)\# 	& -58(299)\# 	& N/A	 
\\
\end{tabular}
\footnotetext{$JYFL - AME16$}
\footnotetext{Measured using both TOF-ICR and PI-ICR techniques, see Table \ref{Eu_results}.}
\end{ruledtabular}
\label{results}
\end{table*}

\subsubsection*{Nuclides measured for the first time: \\$^{161}$Pm, $^{163}$Sm, $^{164,165}$Eu, $^{167}$Gd, $^{165,167,168}$Tb}

The mass-excess values for eight studied nuclides, $^{161}$Pm, $^{163}$Sm, $^{164,165}$Eu, $^{167}$Gd, $^{165,167,168}$Tb, were measured for the first time. The measurements were done with the following excitation patterns in the precision trap: $^{161}\mathrm{Pm}$ (200 ms), $^{163}\mathrm{Sm}$ (200 ms),$^{164}\mathrm{Eu}$ ($\unit[25-750-25]{ms}$ On-Off-On), $^{165}\mathrm{Eu}$ ($\unit[25-350-25]{ms}$ On-Off-On), $^{167}\mathrm{Gd}$ (400 ms), and $^{165-168}\mathrm{Tb}$ ($\unit[25-750-25]{ms}$ On-Off-On). All new mass-excess values agree within 1.5$\sigma$ with the extrapolated value from AME16 \cite{AME16}. The largest deviations to AME16 extrapolation occur for the lightest studied nuclides, $^{161}$Pm, $^{163}$Sm and $^{164}$Eu, all being around 150 keV higher than the AME16 extrapolation (see Fig.~\ref{fig:all-measurements}). 

\subsubsection*{\texorpdfstring{$^{154}$Nd}{}}

\begin{figure*}[tpb]
\begin{center}
\centerline{\includegraphics[trim = 0mm 0mm 0mm 0mm, clip,width=\linewidth]{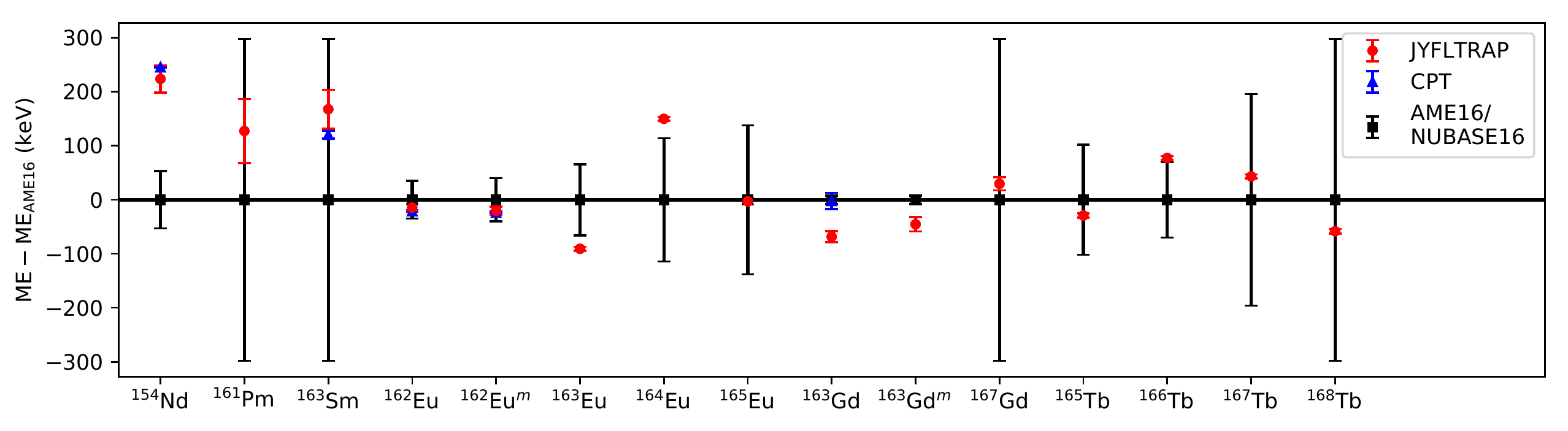}}    	
\caption{(Color online) Comparison to AME16 mass-excess values for each measured nuclide. The results from CPT \cite{vanschelt2012,orford2018,hartley2018} are also shown. The values for the isomeric states were adopted from NUBASE16 \cite{nubase16}.}
\label{fig:all-measurements}
\end{center}
\end{figure*}

The determined mass-excess value of $^{154}$Nd, $-65601(25)$~keV, is based on three consistent individual frequency ratios measured with 200~ms quadrupolar excitation in the precision trap. In AME16, the mass of $^{154}$Nd is based on beta-decay end-point energies of $^{154}$Nd  \cite{Greenwood1993} and $^{154}$Pm \cite{Dauria1971,Tannila1972,Yamamoto1974} connecting the isobaric chain to $^{154}$Sm for which the mass has been directly measured \cite{Kayser1975}. The mass-excess value obtained in this work for $^{154}$Nd is 225 keV higher than the AME16 value \cite{AME16}. This is understandable since the beta-decay studies often suffer from the pandemonium effect \cite{Hardy1977}. It means that transitions from higher-lying states have been missed, leading to too low beta-decay $Q$-values. More recently, CPT has also measured $^{154}$Nd \cite{orford2018}. The JYFLTRAP value agrees well with this recent Penning-trap measurement (see Fig.~\ref{fig:all-measurements}).

\begin{table*}
\caption{Frequency ratios ($r$) and mass-excess values ($ME$) determined in this work for $^{162}$Eu and $^{162}$Eu$^m$ with the PI-ICR and TOF-ICR measurement techniques. The excitation energy $E_x$ for the isomer is also given. The reference mass values were adopted from AME16 \cite{AME16}.}
\begin{ruledtabular}
\begin{tabular}{llllll}
Isotope & Reference & Method	& $r(\delta r_{stat})(\delta r_{tot})$ & $ME(\delta ME_{tot}) (keV)$ & $E_x$ (keV)\\ 
\midrule
$^{162}$Eu			&	$^{162m}$Eu	& PI-ICR		& 0.999 998 966(19) &	-58721.8(3.7) & 155.9(3.0)
\\
			&	$^{133}$Cs		& TOF-ICR		& 1.218 438 235(35)(45) &	-58717.2(4.4)(5.5) & N/A 		
\\
\cmidrule{3-6}
& & Difference &  N/A  & -4.7(6.6) & N/A\\
& & Weighted mean & N/A   & -58720.4(3.1) & N/A\\
\hline
$^{162m}$Eu	&	$^{133}$Cs 		& PI-ICR		& 1.218 439 457(18) &	-58566.0(2.2)	&		N/A
\\
	&	$^{133}$Cs 		& TOF-ICR		& 1.218 439 502(55)(61) &	-58560.3(6.8)(7.6)	&		156.8(9.4)
\\
\cmidrule{3-6}
& & Difference & -0.000 000 045(63)   & -5.6(7.9) & 0.9(9.8)\\
& & Weighted mean & 1.218 439 459(61)   & -58565.7(7.6) & 156.0(2.8)\\
\end{tabular}
\end{ruledtabular}
\label{Eu_results}
\end{table*}

\subsubsection*{\texorpdfstring{$^{162,162m}$Eu}{}}
    
\begin{figure}
\begin{center}
\centerline{\includegraphics[trim = 0mm 0mm 0mm 0mm, clip,width=\linewidth]{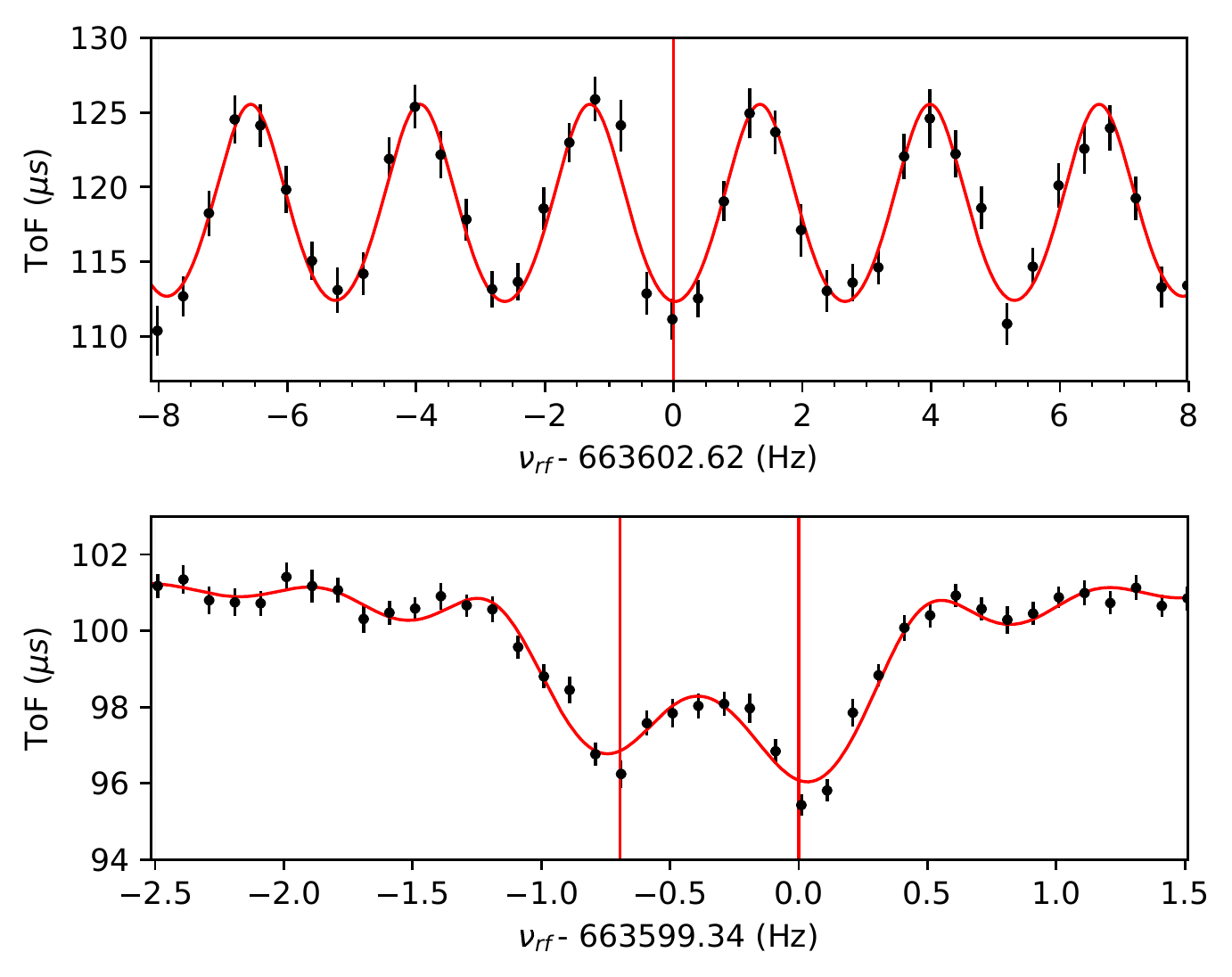}} 
\caption{(Color online) Time-of-flight spectra for $^{162}$Eu$^+$ from campaign I using $\unit[25-350-25]{ms}$ (On-Off-On) excitation pattern (top), and from this work with a 1600~ms excitation (bottom). Fitted theoretical line shapes and $\nu_c$ frequencies are plotted in red.}
\label{fig:162Eu_tof_plot_combined}
\end{center}
\end{figure}

$^{162}$Eu was measured already in the first JYFLTRAP campaign on neutron-rich rare-earth nuclides \cite{vilen2018}. The measurement done with the TOF-ICR method using a Ramsey excitation pattern $\unit[25-350-25]{ms}$ (On-Off-On) resulted in a mass-excess value of $-58658(4)$~keV \cite{vilen2018}. Around the same time, the CPT measured $^{162}$Eu using the PI-ICR method \cite{hartley2018}. They discovered an isomeric state at 160.2(24)~keV above the ground state for which a mass-excess of $-58723.9(15) \mathrm{keV}$ was determined \cite{hartley2018}. The JYFLTRAP mass-excess value for the measured state was around 60 keV above the ground state and 100 keV below the isomer determined at CPT, implying that a mixture of states had been measured at JYFLTRAP. Hence, a remeasurement was performed in this work. It was done both with the TOF-ICR and PI-ICR techniques. For the TOF-ICR measurements, a long, 1600~ms quadrupolar excitation sufficient to separate the two states, was employed. Figure \ref{fig:162Eu_tof_plot_combined} shows TOF-ICR spectra for $^{162}$Eu from the first JYFLTRAP campaign \cite{vilen2018} and from this work. Clearly, the first JYFLTRAP campaign could not have easily identified the isomeric state, unknown at the time, due to the shorter excitation time (leading to worse resolution) and the used Ramsey pattern with several strong minima.

The PI-ICR measurements of $^{162}$Eu and $^{162}$Eu$^m$ were performed as described in Sect.~\ref{sec:pi-icr} with a 600-ms phase accumulation time. As can be seen from Fig.~\ref{162Eu_PI-ICR}, abundances of the two states were similar, with $53.5(1.8)\%$ of the detected ions being in the ground state. This supports the fact that a mixture of states was measured during the first JYFLTRAP campaign. The new TOF-ICR and PI-ICR results for the ground and isomeric state of $^{162}$Eu (see Table~\ref{Eu_results}) are now consistent, confirming that the recently commissioned PI-ICR method works at JYFLTRAP. Both the ground-state and isomeric-state values agree with the CPT results, and with the AME16 value based on a beta-decay study \cite{hayashi2014}. The excitation energy obtained for the isomer at JYFLTRAP is 156.0(2.8) keV  which is somewhat lower than obtained at CPT. 

The weighted means presented in Table~\ref{Eu_results} were calculated between the two measurement techniques. Internal and external uncertainties were calculated and the larger one was accepted as the uncertainty of the mean. In the case of $^{162}\mathrm{Eu}^m$, the larger error of the two individual results was adopted to account for the difference in results between the techniques.

\subsubsection*{\texorpdfstring{$^{163,163m}$Gd}{}, and \texorpdfstring{$^{163}$Eu}{}}

Both $^{163}$Gd and $^{163}$Eu were measured already in the previous campaign at JYFLTRAP \cite{vilen2018}. There, Ramsey excitation with a pattern $\unit[25-350-25]{ms}$ (On-Off-On) was used together with a reference from the same isobaric mass chain, $^{163}$Dy. The resulting $^{163}$Gd mass-excess value was $-61200(4)$~keV. Surprisingly, this deviated from the AME16 value, based on a CPT measurement \cite{vanschelt2012}, by 114(9)~keV. Since the difference was very close to the first isomeric state energy, 137.8 keV \cite{hayashi2014}, it was suggested that JYFLTRAP had measured the isomer, which could be predominantly populated in the proton-induced fission of $^{238}$U used at IGISOL. The spontaneous fission of $^{252}$Cf used for the production at CPT could, in turn, populate predominantly the ground state. 

To confirm whether an isomeric state had been measured in \cite{vilen2018}, $^{163}$Gd was remeasured at JYFLTRAP using the TOF-ICR technique with a long, 1600~ms quadrupolar excitation, which is sufficient to separate the two states. In addition, a more accurate reference, $^{136}$Xe, was used. A total of five consistent frequency ratio measurements were done, yielding a ground-state mass-excess of $-61382.4(10.2)$~keV, and an excitation energy of $161(17)$~keV for the isomer. The new value still disagrees with the CPT value ($-61316.0(15.0)$ \cite{vanschelt2012}) and is not consistent with the excitation energy ($137.8$~keV) from \cite{hayashi2014}. 

Counter to the hypothesis, the re-measurement showed that the ground state was dominant. Based on a PI-ICR yield measurement of $^{163}$Gd, the yield ratio was $72.5(5.0)\%$ for the ground state at the MCP detector. Hence, the first $^{163}$Gd measurement at JYFLTRAP should have produced a mass-excess value close to the ground-state value obtained in the re-measurement. The 180~keV difference between the two $^{163}$Gd measurements casts doubt on the reference used in the first measurement, $^{163}$Dy. This reference was chosen at the time because the mass-excess value of $^{163}$Dy is well known, with a precision of 0.8~keV \cite{AME16}, based on a recent Penning-trap measurement at TRIGA-TRAP \cite{Schneider2015}. Using a reference from the same isobaric mass chain is also preferential since it cancels out any mass-dependent uncertainties. Since the $^{163}$Eu measurement from \cite{vilen2018} had used the same reference, it was decided to remeasure also $^{163}$Eu, this time with an unambiguous reference, $^{133}$Cs$^+$, from the off-line ion source station. A total of five consistent frequency ratios with a Ramsey excitation pattern of $\unit[25-350-25]{ms}$ (On-Off-On) were performed, resulting in a mass-excess value of $-56575.7(3.8)$~keV. This differs by 156 keV from the previous value, $-56420(4)$~keV, reported in \cite{vilen2018}. The similar systematic shift as for the remeasured $^{163}$Gd confirms the suspicion that the $^{163}$Dy reference had been wrongly identified.

In order to correctly identify the $A= 163$ reference ion used in the first campaign \cite{vilen2018}, we used the software SCM$\_$Qt (Search for Contaminant Masses) \cite{ringleThesis2006} which lists all possible molecular combinations of a specified list of elements and maximum number of atoms of those elements that would yield the same frequency ratio. For the calculations, we used the average cyclotron frequency of $^{171}$Yb$^+$ (628665.80 Hz) as a reference since it was the closest reference measurement before the use of the to-be-identified $A$ = 163 ion (659564.18 Hz). We allowed SCM to form molecules of up to three different elements and up to 10 atoms of the same elements chosen from H, C, O, N, and fission products. We also only kept the isotopes with a half-life greater than 100 ms. As a result, the closest reasonable candidate found was $^{146}$La$^{16}$O$^{1}$H$^+$. $^{146}$La is produced with a large cross section in proton-induced fission, with an estimated rate of around 6.5$\times$10$^{4}$ particles/s. 

When using $^{146}$La$^{16}$O$^{1}$H$^+$ as reference for the initial measurement of $^{163}$Eu \cite{vilen2018}, we obtain a mass-excess value of $-56539(37)$~keV. Most of the uncertainty stems from the 30 keV uncertainty in the mass excess of $^{146}$La, which is based on a Penning-trap measurement at CPT \cite{Savard2006} and several $\beta$-decay end-point energy measurements \cite{AME16}. This result is within one standard deviation from the new measurement presented in Table~\ref{tab:results}. 

Similarly, using $^{146}$La$^{16}$O$^{1}$H$^+$ as a reference for the initial measurement of $^{163}$Gd \cite{vilen2018}, we obtain a mass-excess value of -61319(37) keV, which is 1.7$\sigma$ higher than the new measurement presented in Table~\ref{tab:results}. Such a discrepancy is not unexpected, since the initial measurement from \cite{vilen2018} was performed using a Ramsey pattern $\unit[25-350-25]{ms}$ (On-Off-On), which is not sufficient to distinguish between the ground and isomeric states. As such, the measured mixture of states in \cite{vilen2018} is expected to produce a value in between the two states as in the case of $^{162}$Eu. The CPT measurement from \cite{vanschelt2012}, which was done with a too short excitation time (500 ms) to resolve the two states, yields a similar mass-excess value of $-61316(15)$ keV. It should also be noted that the frequency difference between $^{146}$La$^{16}$O$^{1}$H$^+$ and $^{163}$Dy$^+$ is less than 1 Hz in JYFLTRAP, and hence, those cannot be resolved in the purification trap and would have been difficult to identify with the used Ramsey-type excitation in the precision trap. Therefore, it is possible that the reference was a mixture of $^{146}$La$^{16}$O$^{1}$H$^+$ and $^{163}$Dy$^+$.

\subsubsection*{$^{166}$Tb}

The mass of $^{166}$Tb was determined based on six individual frequency ratio measurements done with a Ramsey excitation pattern $\unit[25-750-25]{ms}$ (On-Off-On). The resulting mass-excess value is 77(71)~keV above the AME16 value stemming from a $\beta$-decay endpoint energy measurement \cite{HAYASHI2010}. The precision of the mass-excess value was improved by a factor of almost 20.

\section{Impact on the mass surface and its derivatives \label{sec:surface}}

By plotting atomic masses as a function of neutron ($N$) and proton ($Z$) numbers, a surface in a three-dimensional space is obtained. If we neglect the pairing effect, i.e. by selecting only even-even, odd-odd, odd-even or even-odd nuclei, the surface is rather regular and smooth. Sudden changes or irregularities on the surface can be caused e.g. by shell closures at magic nucleon numbers or changes in the shape of the ground state. To reveal possible changes in nuclear structure and interactions, it is very useful to study different derivatives of the mass surface, such as one- and two-neutron separation energies ($S_n$ and $S_{2n}$), neutron pairing gap energies $D_n$, two-neutron shell gap energies $D_{2n}$, and proton-neutron pairing strength metrics $\delta V_{pn}$. In the following sections, the data presented in Table~\ref{tab:results} were used to study these mass derivatives. Each of the studied quantities is sensitive to some aspect of nuclear structure and offers valuable feedback for theoretical mass models needed e.g. for the unknown neutron-rich nuclei in the abundance calculations of the $r$ process.

\subsection{Neutron separation energies \label{sec:Sn}}

Neutron separation energies are one of the most influential inputs needed for the $r$-process modeling, and indeed the primary means by which mass measurements influence the calculated rare-earth abundances in the $r$ process. The neutron separation energy, $S_n$, is determined as
\begin{equation}\label{eq:Sn}
S_n(Z,N) = [m(Z,N-1) + m_n - m(Z,N)]c^2\text{,}
\end{equation}
where $m$ denotes the masses for the nuclides $(Z,N)$, $(Z,N-1)$ and the neutron, and $c$ is the speed of light in vacuum. The neutron separation energy appears directly in neutron-capture rates, which for radioactive nuclides of the type presented, are largely unmeasured, and therefore must be calculated via statistical models \cite{mumpower2015prc035807}. It also appears exponentially in photo-dissociation rates, which are perhaps the most important factors shaping the $r$-process path, highlighting the direct impact mass measurements have on $r$-process calculations.

\begin{figure}
	\begin{center}
    	\centerline{\includegraphics[trim = 20mm 10mm 60mm 20mm, clip,width=\linewidth]{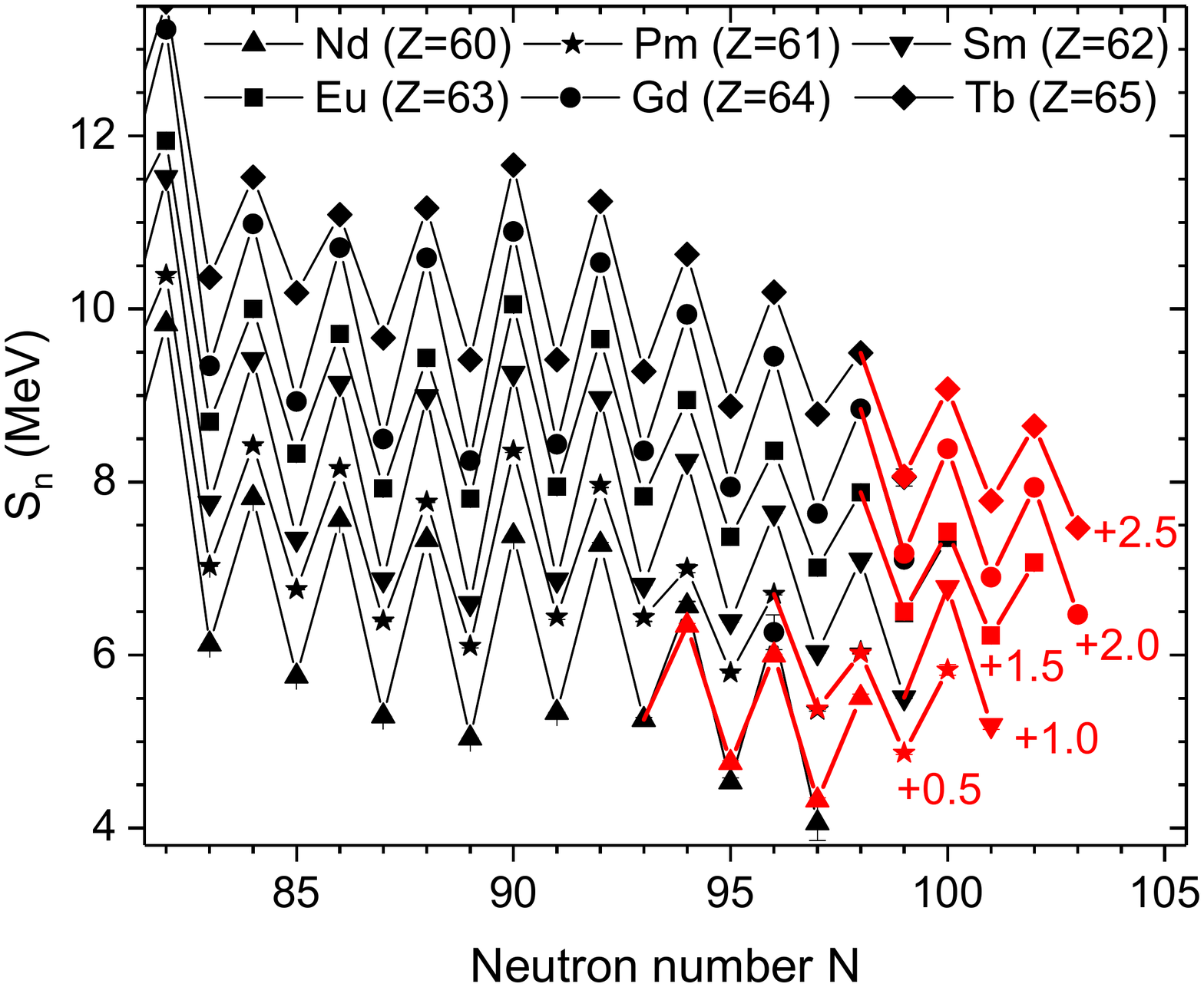}}    	
			\caption{Experimental neutron separation energies, $S_n$, from  AME16 \cite{AME16} (in black), and supplemented with the JYFLTRAP results from \cite{vilen2018} and this work (in red) . Each isotopic chain has been shifted by $\unit[0.5]{MeV}$ relative to the previous chain for clarity, with the Nd chain remaining unchanged.}
    \label{fig:Sn-2}
	\end{center}
\end{figure} 

\begin{figure}
	\begin{center}
    	\centerline{\includegraphics[trim = 0mm 0mm 0mm 0mm, clip,width=\linewidth]{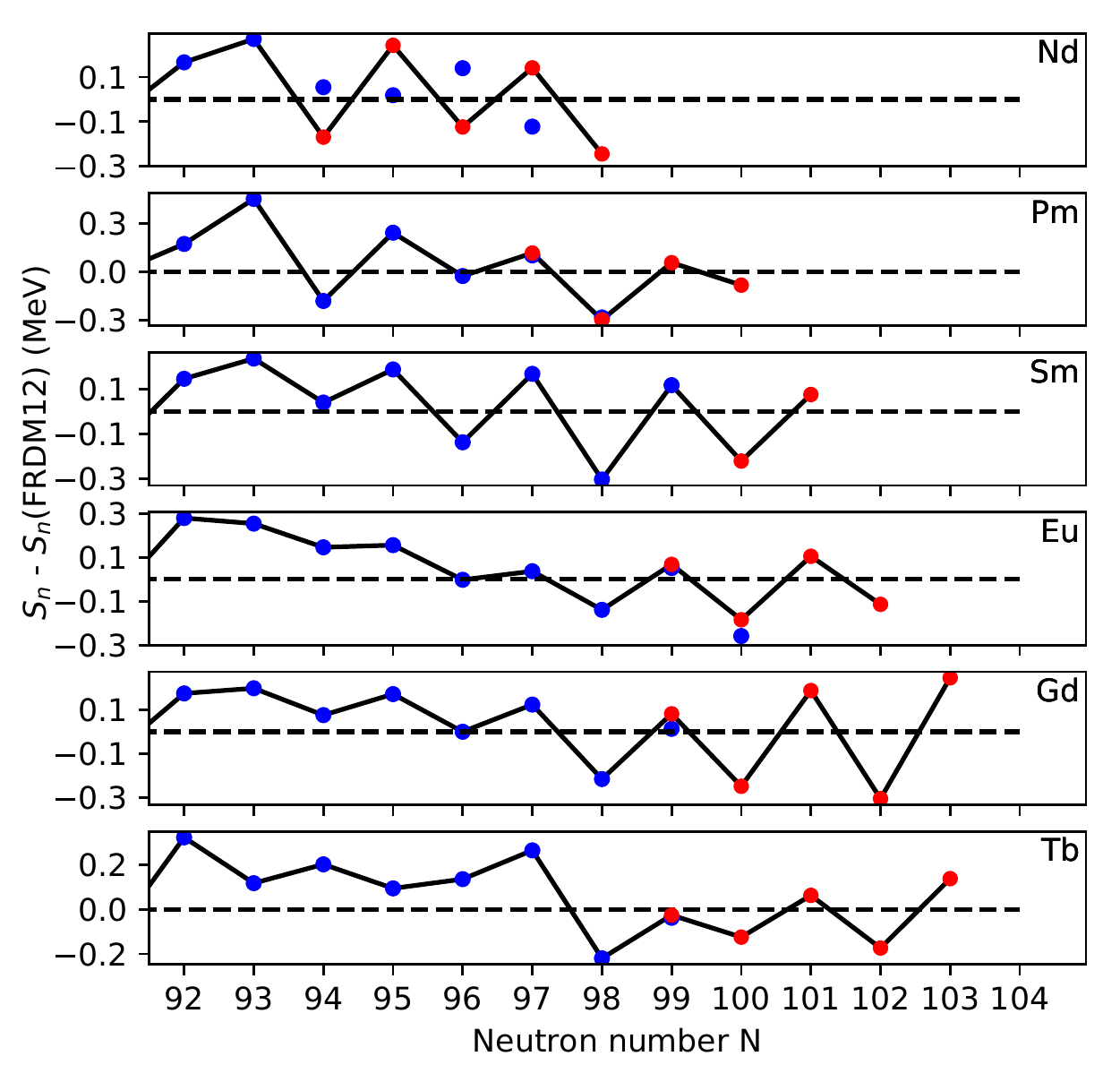}}    	
			\caption{Experimental neutron separation energies $S_n$ compared to the FRDM12 mass model. The $S_n$ values affected by the JYFLTRAP measurements of this work and \cite{vilen2018} are shown in red and AME16 values in blue. Dashed line indicates energies from FRDM12.}
    \label{fig:Sn_differences}
	\end{center}
\end{figure}

Figure~\ref{fig:Sn-2} shows the experimental neutron separation energies for isotopic chains from the element neodymium ($Z = 60$) to terbium ($Z = 65$) in the region involving our new measurements. When comparing to AME16 \cite{AME16}, the new $S_n$ values do not immediately reveal any radical changes to the previously known trends. However, the new neodymium measurements supplant four existing, successive literature values from $N = 94$ to $N = 97$, and a notable reduction in odd-even staggering is clearly seen. Similarly to the high precision measurements of Nd, which show a decrease in odd-even staggering relative to existing experimental data, there is also a tendency for the theoretical models to overestimate this effect. This is illustrated in Fig.~\ref{fig:Sn_differences} for FRDM12 \cite{Moller2016}. The new experimental $S_n$ values are higher (lower) than in FRDM12 for odd (even) $N$, decreasing the odd-even staggering for the most neutron-rich isotopes. 

\begin{table}
\caption{Root mean square errors, $\delta_{RMS}$ (see Equation \ref{eq:rms}), between the measured and theoretical $S_n$ values, as well as the over-prediction of the neutron pairing gap, $\delta D_n$ (see Equation \ref{eq:delta-Dn}) over $60 \leq Z \leq 65$ with $N\geq94$.} 
\begin{ruledtabular}
\begin{tabular}{lll}
Mass Model & $\delta_{RMS}$ (MeV) & $\delta D_n$ (MeV)	 \\ 
\midrule
HFB-24 & 0.356 & 26.9  \\
FRDM12 & 0.168 & 11.2  \\
Duflo-Zuker & 0.141 & 6.5  \\
WS3 & 0.186 & 13.9 \\
WS3+ & 0.224 & 16.4  \\
WS4 & 0.190 & 14.4 \\
WS4+ & 0.222 & 16.8  \\
\end{tabular}
\end{ruledtabular}
\label{tbl:SnRMSerrors}
\end{table}

In addition to FRDM12 \cite{Moller2016}, experimental $S_n$ values were compared to other theoretical mass models typically used in astrophysical $r$ process calculations, including HFB24 \cite{Goriely2013}, Duflo-Zuker \cite{duflo1995}, WS3 \cite{WS3wang2013}, WS3+ \cite{WS3+wang2011}, WS4 \cite{Wang2014}, and WS4+ \cite{Wang2014}. In order to facilitate the comparison, root-mean-square (RMS) errors were calculated for the models according to
\begin{equation}
\delta_{RMS} = \sqrt{\frac{1}{N_{tot}}\sum\limits_{Z,N} \left(S_n(Z,N)_{th.} - S_n(Z,N)_{exp.}\right)^{2}}\text{,}
\label{eq:rms}
\end{equation}
where $N_{tot}$ is the total number of isotopes in the chain, and the sum is over the differences between the theoretical ($th.$) and experimental ($exp.$) values of $S_n$ across all isotopes of each chain, and $Z$ and $N$ are the proton and neutron numbers, respectively. Table~\ref{tbl:SnRMSerrors} indicates that the calculated RMS errors range from 141 to 356 keV, with any given isotopic chain being within $\delta_{RMS}=483$~keV from the JYFLTRAP values. Although the $S_n$ values from FRDM12  deviated from the experimental values more than Duflo-Zuker in this mass region, FRDM12 still remains a benchmark against which to compare our measurements. As discussed in \cite{Mumpower2017}, this is because $r$-process simulations using FRDM12 yield isotopic abundances that more closely match the solar data in this region.

\subsection{Neutron pairing gaps \label{sec:Dn}}
	       
\begin{figure}[tpb]
\begin{center}
\centerline{\includegraphics[trim = 15mm 10mm 30mm 10mm, clip,width=\linewidth]{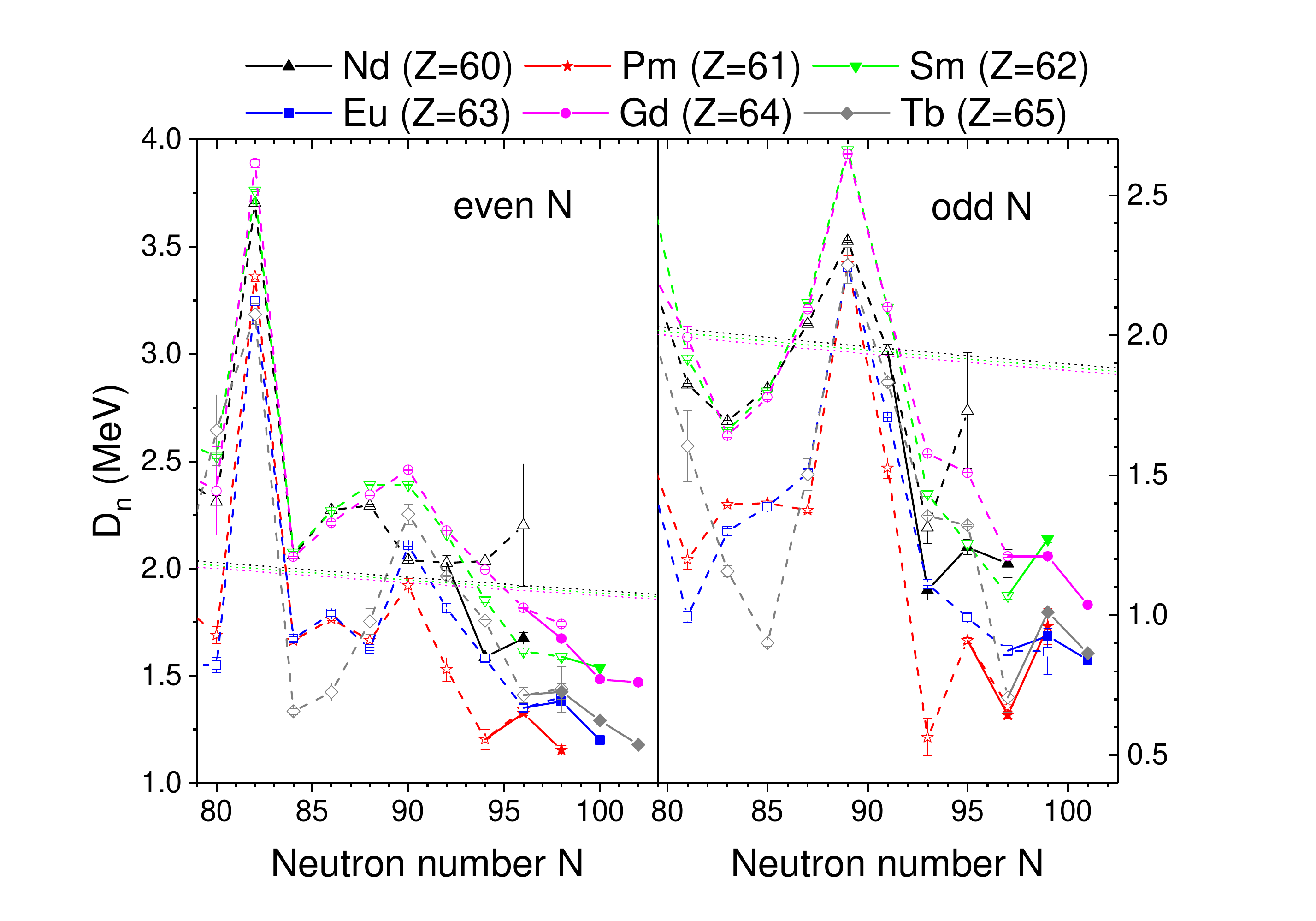}}    	
\caption{Pairing-gap energies $D_n$ for even-$N$ (left) and odd-$N$ (right) isotopes of the studied isotopic chains based on the experimental AME16 \cite{AME16} mass values (dashed lines, open symbols) and the JYFLTRAP measurements from this work and \cite{vilen2018} (solid thick lines, full symbols). The dotted lines present $D_n$ for $Z=60$, $62$ and $64$ based on the average parametrization of the pairing gap energy $\Delta\approx 12/\sqrt{A}$~MeV \cite{Bohr1998}. Note the different scales for even and odd $N$.}
\label{fig:Dn}
\end{center}
\end{figure}

The odd-even staggering of neutron separation energies is a strong signature of neutron pairing. The neutron pairing can be quantified in terms of the differences between the neutron separation energies $S_n$ of successive isotopes: 
\begin{equation}\label{eq:Dn}
D_n(Z,N)=(-1)^{N+1}\big[S_n(Z,N+1)-S_n(Z,N)].
\end{equation}
This neutron pairing energy metric $D_n$ \cite{brown2013} is twice the odd-even staggering parameter $\Delta^{(3}_n(N)$, also commonly used to describe the neutron pairing gap, see e.g. \cite{hakala2012}. Figure~\ref{fig:Dn} shows the $D_n$ values for the various isotopic chains affected by our new atomic mass measurements. The odd-$N$ and even-$N$ data are plotted separately due to different factors affecting the values. The odd-$N$ values can be directly associated with the pairing effect while even-$N$ contain also contribution from the single-particle splitting around the Fermi level \cite{Satula1998,Duguet2001}. The peak at the closed neutron shell at $N$ = 82 is prominent in the even-$N$ $D_n$ values. The onset of deformation at around $N=89$ is manifested as a peak both in the even-$N$ and odd-$N$ $D_n$ values, followed by a gradual decrease toward the neutron midshell at $N=104$. The pairing is reduced faster than predicted by the average parametrization \cite{Bohr1998} (see Fig.~\ref{fig:Dn}). The general trend is similar to \cite{Litvinov2005} where the odd-even staggering was found to increase towards the proton dripline.

Previously, a sub-shell gap has been predicted at $N$ = 100 \cite{Ghorui2012,Satpathy2003,Satpathy2004}. With our new mass measurements going beyond the $N=100$ closure (see Fig.~\ref{fig:Dn}), no unusual increase in the $D_n$ values (as seen at $N$ = 82) is present at $N$ = 100, negating the presence of a sub-shell closure or onset of deformation (as seen at $N$ = 89). 
            
\begin{figure}[tpb]
\begin{center}
\centerline{\includegraphics[trim =15mm 10mm 35mm 20mm, clip,width=\linewidth]{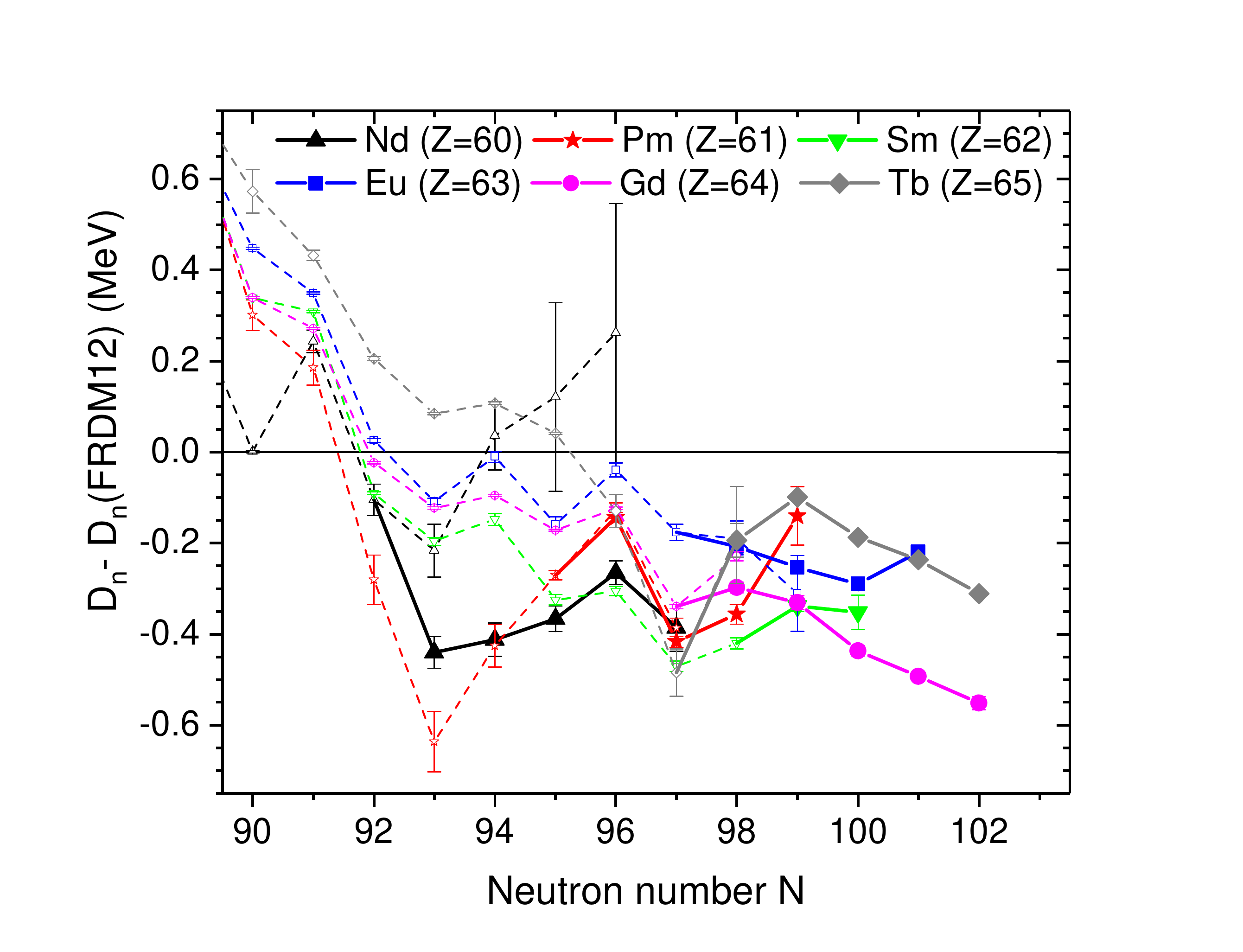}} 
\caption{Differences in pairing-gap energies, $D_n$, between the experimental values taken from AME16 \cite{AME16} (dashed lines, open symbols) and the JYFLTRAP measurements from this work and \cite{vilen2018} (thick solid lines, full symbols), and the FRDM12 mass model \cite{Moller2012} (baseline). JYFLTRAP measurements extend the AME16 trend in showing that FRDM12 consistently over-estimates the effect of pairing for neutron-rich nuclei.}
\label{fig:Dn-diff}
\end{center}
\end{figure}

To see the effect of the new measurements on the pairing gaps relative to theoretical mass models, Fig.~\ref{fig:Dn-diff} shows the difference in the $D_n(N)$ values between the FRDM12 mass model \cite{Moller2012} (which tends to be more accurate than other models at predicting rare-earth abundances that match solar data) and the AME16 \cite{AME16}, supplemented by the previous \cite{vilen2018} and new JYFLTRAP mass measurements. Relative to FRDM12, a consistent trend of smaller pairing gaps for increasing $N$ emerges (negative values in the plot). This stems from the measured nuclei being generally less tightly bound, and translates to the effect of pairing being weaker than FRDM12 predicts. 
            
Finally, we investigated whether the over-estimation in the $D_n$ values near $N = 100$ is also present in other mass models by calculating the following metric:
\begin{equation}
\delta D_n = \sum\limits_{Z,N} \left(D_n(Z,N)_{th.} - D_n(Z,N)_{exp.}\right).
\label{eq:delta-Dn}
\end{equation}
The $\delta D_n$ values are given for commonly used mass models in Table \ref{tbl:SnRMSerrors}. They show that FRDM12's overprediction of the $D_n$ is not an isolated case since most mass models tend to overpredict this quantity for the studied mass region, $N\geq94$. The overprediction of neutron pairing energies by Duflo-Zuker was the lowest of the models examined, probably because of its featureless, uniform odd-even staggering across the measured isotopic chains. Meanwhile, HFB-24, the strongest over-predictor of $D_n$, is consistently high relative to the values calculated using AME16 even for $N<94$.

\subsection{Two-neutron separation energies \label{sec:S2n}}
        
Changes in nuclear structure far from stability can be probed via two-neutron separation energies \cite{LunneyMasses}, $S_{2n}$,:
\begin{equation}\label{eq:S2n}
S_{2n}(Z,N)=[m(Z,N-2)+2m_n-m(Z,N)]c^2.
\end{equation}
The main advantage of looking at $S_{2n}$ values is that it removes the large odd-even staggering seen in the $S_n$ values, resulting in a smoother pattern that will highlight features that would have been hidden by the effect of pairing.
        
\begin{figure}\begin{center}
\centerline{\includegraphics[trim = 15mm 10mm 35mm 20mm, clip,width=\linewidth]{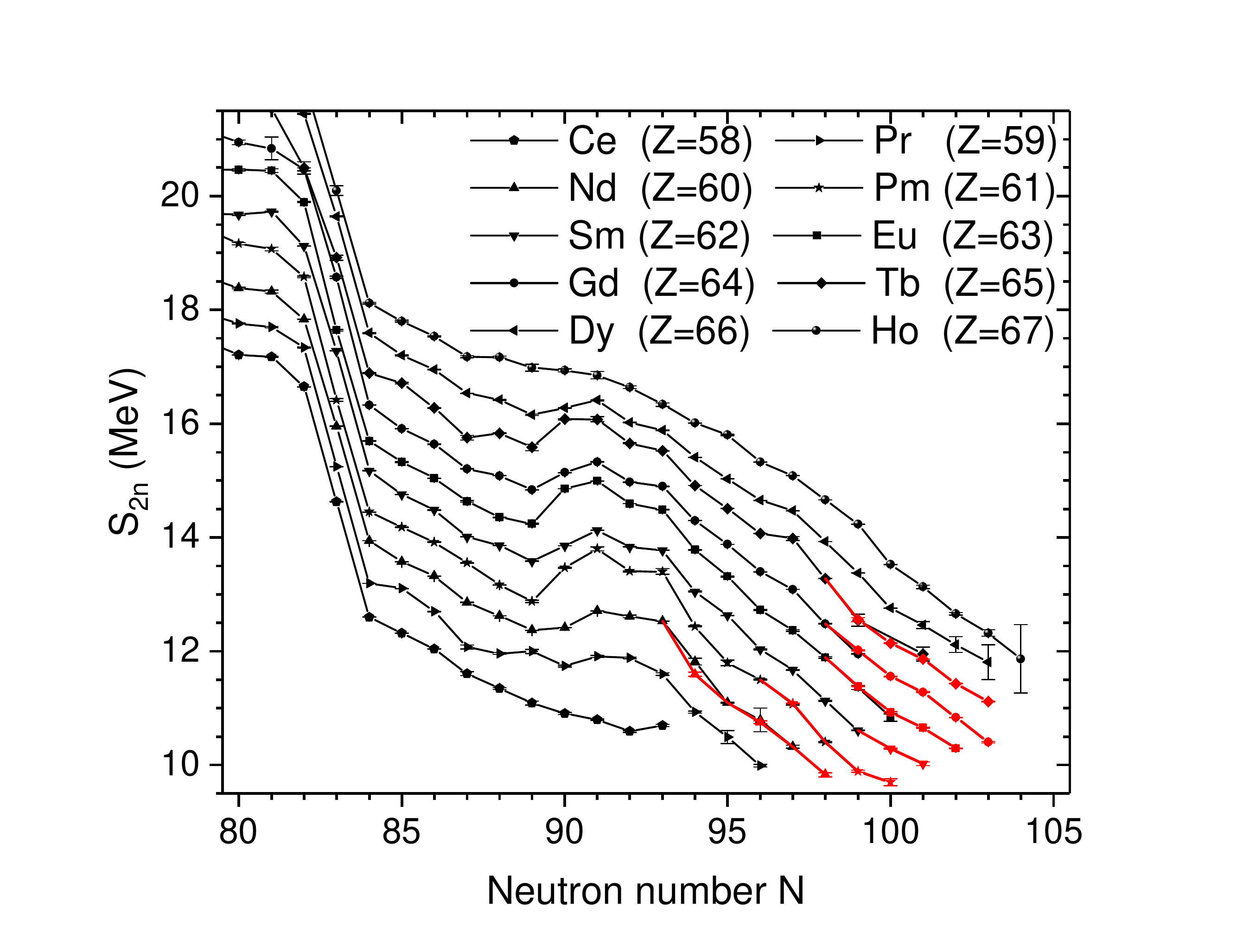}}    	
\caption{Two-neutron separation energies, $S_{2n}$, based on the AME16 \cite{AME16} experimental mass values (in  black) and the JYFLTRAP mass measurements from this work and \cite{vilen2018} (in red).}
\label{fig:S2n}
\end{center}\end{figure}

Two-neutron separation energies for the neutron-rich rare-earth nuclides are plotted in Fig.~\ref{fig:S2n}. The magic shell closure at $N=82$ is seen as a sudden drop in the $S_{2n}$ values whereas the onset of strong prolate deformation at $N$ = 89 produces a kink in the energies. On the other hand, the $S_{2n}$ values in the region affected by our measurements follow a rather smooth behavior except for a sudden change in the slope at $N$ = 97, strongly present only in the Tb chain. This effect will be investigated further by calculating differences in the $S_{2n}$ values.

\subsection{Two-Neutron shell gap energies \label{sec:shellgap}}
          
\begin{figure}[tpb]
\begin{center}
\centerline{\includegraphics[trim = 25mm 10mm 40mm 20mm, clip,width=\linewidth]{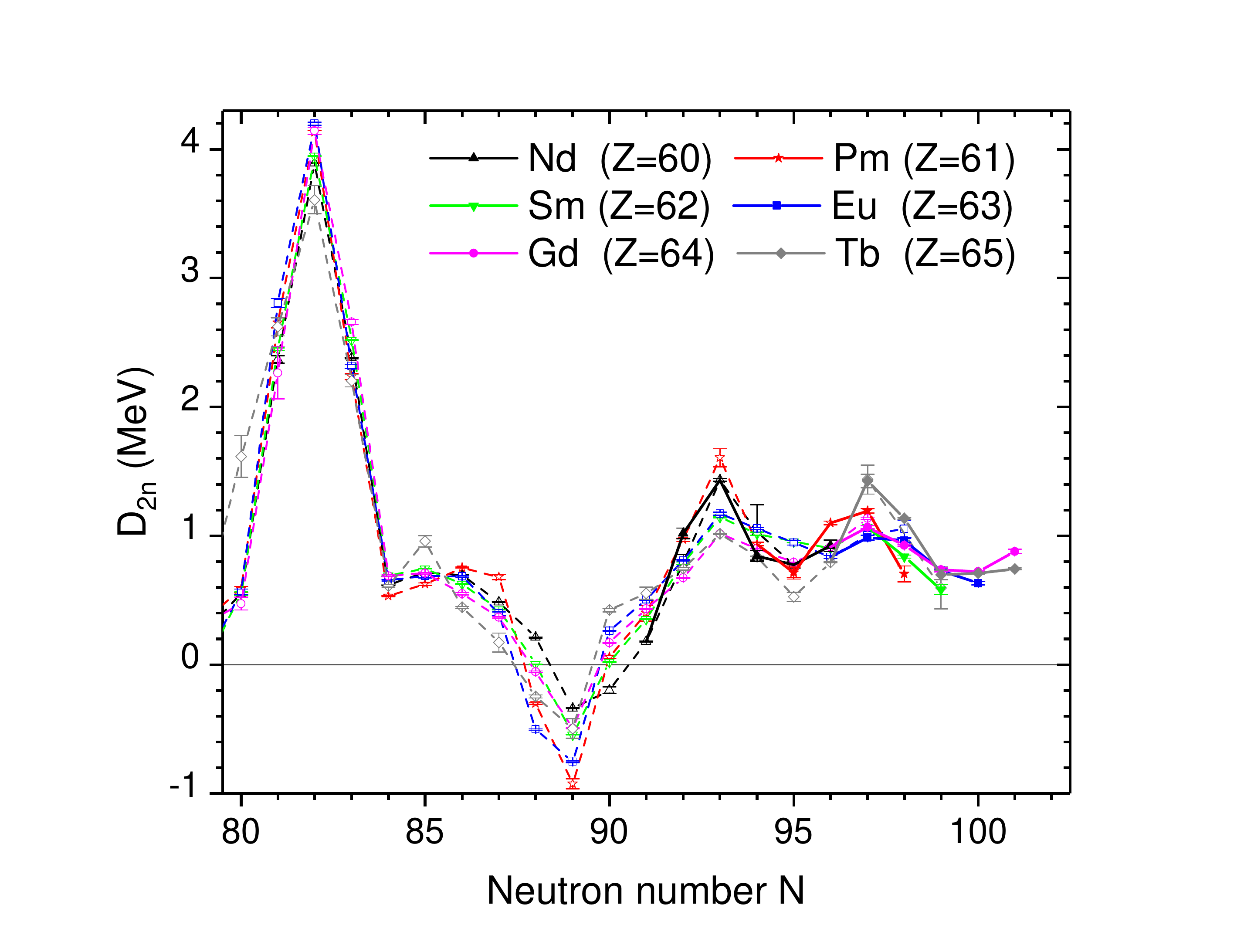}} 
\caption{$D_{2n}$ values based on the AME16 \cite{AME16} experimental mass values (dashed lines and open symbols) and the JYFLTRAP mass measurements from this work and \cite{vilen2018} (thick solid lines and full symbols). No subshell at $N=100$ is apparent but two peculiar peaks appear at $N=93$ and $N=97$.}
\label{fig:D2n}
\end{center}
\end{figure} 

Changes in the slope of $S_{2n}$ values with the number of neutrons can be investigated by calculating the two-neutron shell-gap energies, $D_{2n}$:
\begin{equation}\label{eq:D2n}
D_{2n}(Z,N)=S_{2n}(Z,N)-S_{2n}(Z,N+2).
\end{equation}

Figure~\ref{fig:D2n} shows the $D_{2n}$ values for all isotopic chains affected by our measurements. At magic shell closures, such as at $N=82$, the two-neutron shell-gap energies increase rapidly. At the proposed subshell gap at $N=100$, no increase is observed, contrary to the $2^+$ energies (see Fig.~\ref{fig:e2e4}). Hence, the mass measurement data do not support such a subshell closure. It must be noted, however, that based on the $2^+$ and $4^+$ excitation energies (see Fig.~\ref{fig:e2e4}), the nuclei in this region are not spherical and deformation can affect the two-neutron separation energies, and consequently the shell-gap energies.

\begin{figure}[tpb]
\begin{center}
\centerline{\includegraphics[trim = 20mm 10mm 35mm 20mm, clip,width=\linewidth]{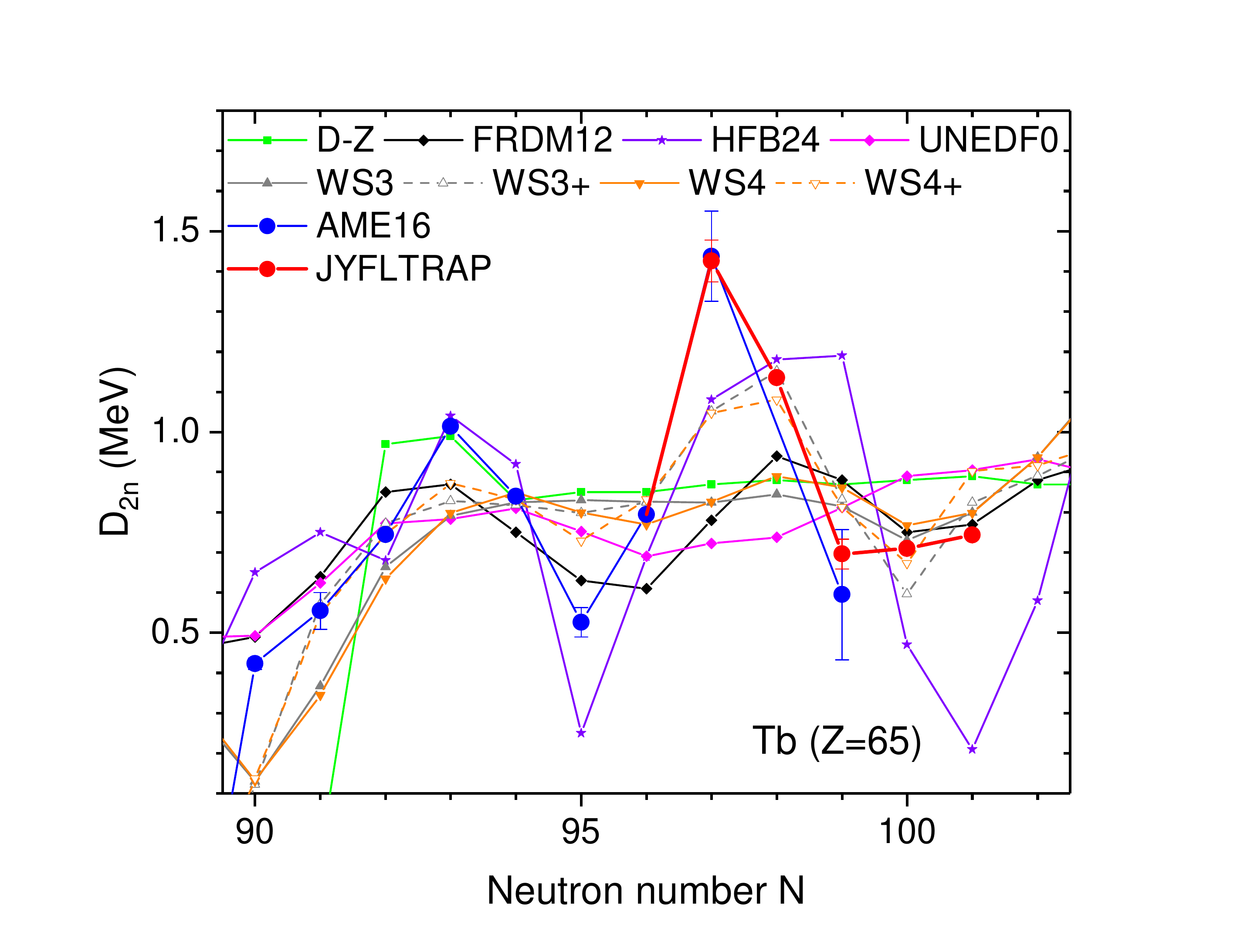}} \caption{Experimental $D_{2n}$ values for the Tb isotopic chain, based on AME16 \cite{AME16} (in blue) and the JYFLTRAP measurements from this work and \cite{vilen2018} (in red), and comparison to various commonly used mass models, such as Duflo-Zuker (D-Z) \cite{duflo1995}, FRDM12 \cite{Moller2016}, HFB24 \cite{Goriely2013}, UNEDF0 \cite{Kortelainen2010}, WS3 \cite{WS3wang2013}, WS3+ \cite{WS3+wang2011}, WS4 \cite{Wang2014}, and WS4+ \cite{Wang2014}. None of these models predict the enhancement seen at $N$ = 97.}
\label{fig:D2n-Tb-Models}
\end{center}
\end{figure}

Figure~\ref{fig:D2n} indicates the presence of two smaller peaks at $N$ = 93 and 97. These peaks are particularly interesting as they are strongest for two odd-odd nuclei, $^{154}$Pm and $^{162}$Tb. Typically such an enhancement is expected only for even-even nuclei. Interestingly, both $^{154}$Pm and $^{162}$Tb have an equal number of valence nucleons above the magic shell closures at $Z=50$ and $N=82$, 11 and 15, respectively. This feature was further investigated by comparing the experimental $D_{2n}$ values with various mass models, such as FRDM12 \cite{Moller2012}, Duflo-Zuker \cite{duflo1995}, and HFB-24 \cite{Goriely2013}, for the Tb chain. Figure~\ref{fig:D2n-Tb-Models} shows that none of these models predicts the observed unusual peak at $N$=97.

\subsection{Proton-neutron pairing strength via the \texorpdfstring{$\delta V_{pn}$}{} values}\label{sec:dVpn}

Neutron-rich rare-earth nuclei with equal numbers of valence protons and neutrons above the closed proton and neutron shells at $Z=50$ and $N=82$ have been proposed to present enhanced proton-neutron interactions \cite{Bonatsos2013}. This is in analogy with the nuclei close to the $Z=N$ line, which exhibit peaks in the proton-neutron interactions for nuclides with maximal spatial-spin overlaps of proton and neutron wave functions \cite{talmi1962,vanisacker1995}. Here we want to investigate whether the unusually large $D_{2n}$ values observed for odd-odd nuclei with $Z_{val}=N_{val}=11$ and $Z_{val}=N_{val}=15$ at $N=93$ and $97$, respectively, could be explained by proton-neutron interactions.

A simple way to study the proton-neutron interaction is to calculate a double difference of binding energies across isotopic chains, called $\delta V_{pn}$. This quantity is defined for odd-odd nuclei as
\begin{equation}\label{eq:dVpn}
\delta V_{pn}(Z_{odd},N_{odd})=S_n(Z,N)-S_n(Z-1,N).
\end{equation}
The odd-odd nuclei are often chosen because their last protons and neutrons occupy specific single orbits, and therefore offer a clearer perspective on the valence $p-n$ interactions.

\begin{figure}[tpb]
\begin{center}
\centerline{\includegraphics[trim = 20mm 10mm 40mm 20mm, clip,width=1\linewidth]{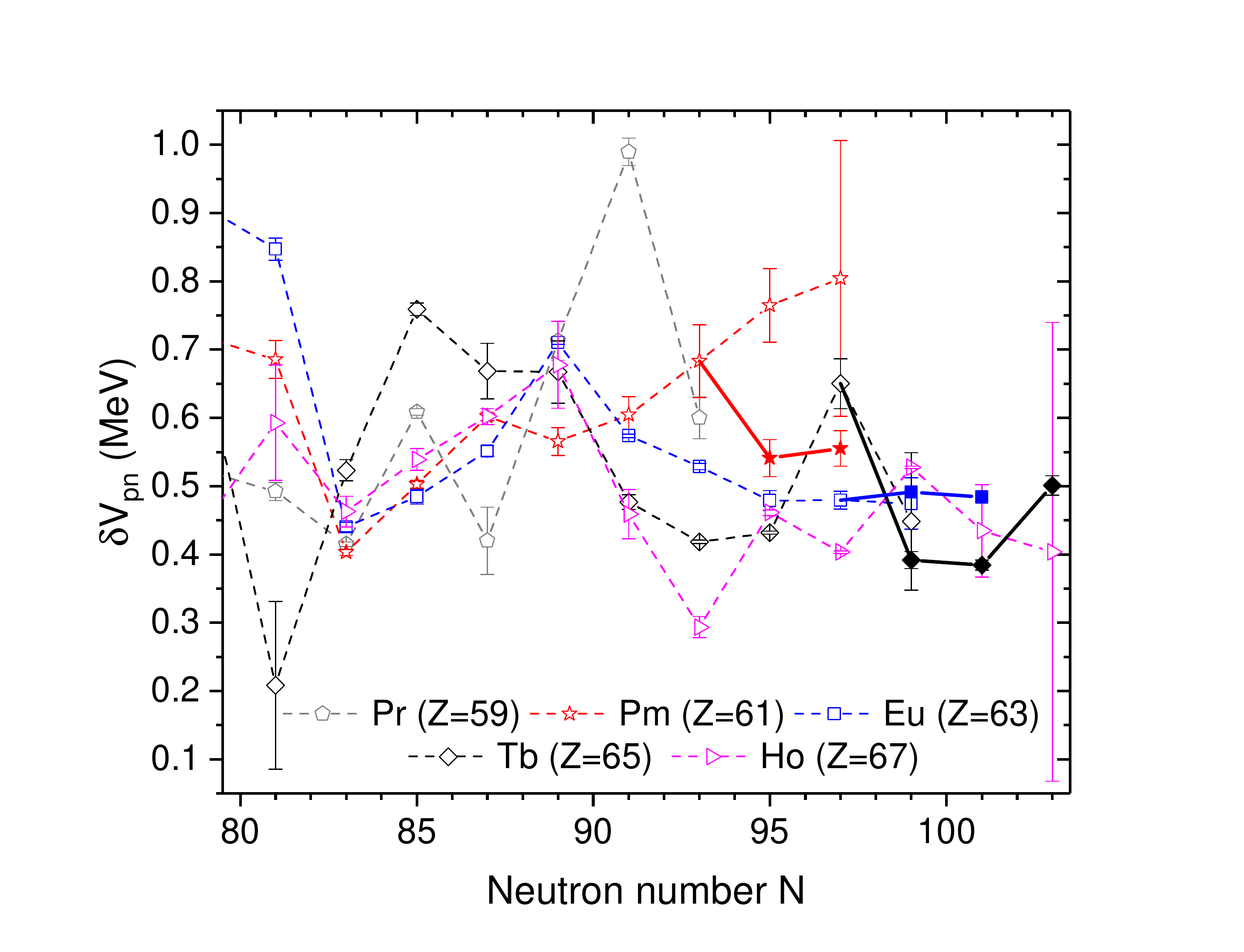}}    	
\caption{A measure of the average $p-n$ interaction of valence nucleons in odd-odd nuclei according to Eq.~\ref{eq:dVpn}. Calculations were done using the experimental mass values from AME16 \cite{AME16} (dashed lines and open symbols) and supplemented by the JYFLTRAP measurements from this work and \cite{vilen2018} (solid lines and full symbols).}
\label{fig:dVpn_oddodd}
\end{center}
\end{figure} 

\begin{figure}[tpb]
\begin{center}
\centerline{\includegraphics[trim = 25mm 10mm 35mm 20mm, clip,width=1\linewidth]{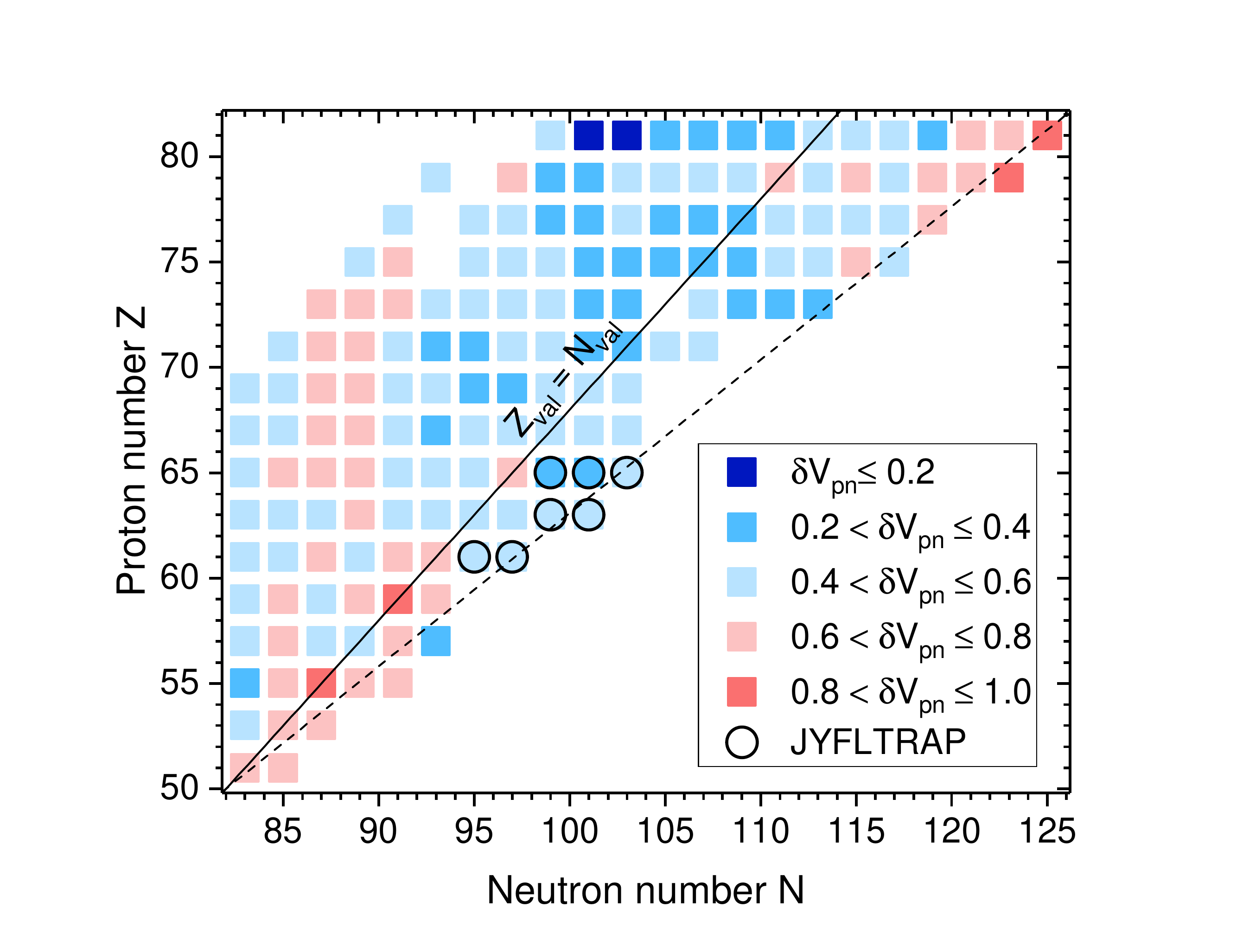}}    	
\caption{Odd-odd $\delta V_{pn}$ values (in MeV) for $80 \leq N \leq 126$ and $50 \leq Z \leq 82$ according to Eq.~\ref{eq:dVpn}. Values affected by JYFLTRAP measurements are circled, and black lines indicate $Z_{val}=N_{val}$ (solid) and equal fractions of the $N$ and $Z$ shells filled (dashed).}
\label{fig:dVpn_oddodd-chart}
\end{center}
\end{figure}

 Figure~\ref{fig:dVpn_oddodd} shows that indeed our new mass measurements unveil the presence of local maxima in the $\delta V_{pn}$ values for the Pm isotopes at $N=93$ ($Z_{val}=N_{val}=11$) and for the Tb chain at $N=97$ ($Z_{val}=N_{val}=15$). Hence, the increased proton-neutron pairing could be an explanation also for the peaks observed in the $D_{2n}$ values. The new measurements support the conclusion that there is enhanced proton-neutron pairing in nuclei with $Z_{val}=N_{val}$ (see Fig.~\ref{fig:dVpn_oddodd-chart}). Interestingly, for La and Eu isotopes the effect is not observed using the present mass measurement data.

\section{Impact on the rare-earth abundance peak}\label{sec:rprocessimpacts}

The rare-earth abundance peak at around $A=165$ gives us essential information on the $r$ process and related nuclear structure effects. Namely, deformation in this midshell region close to $Z=66$, $N=104$ can funnel the reaction flow toward $A\approx 165$ when matter is decaying toward stability at later stages of the process when it is running out of neutrons \cite{Surman1997,mumpower2012_RE}. On the other hand, these lanthanide nuclei can also be populated via fission cycling from asymmetric fission of heavy nuclei in the $A\approx 280$ region \cite{Goriely2013}. The lanthanides play a key role in the emergence of the red kilonova after GW170817 \cite{Kasen2017,Smartt2017} due to their much higher opacity than lighter elements. In order to provide accurate $r$-process calculations to interpret multimessenger observational data from neutron star mergers, it is essential to have precise mass values of the lanthanide nuclides in the rare-earth region. It should be noted that the observed solar system abundances of lanthanides are one of the most well known \cite{Lodders2009}. Europium, for instance, is the standard representative element for the $r$-process in stellar observations (see e.g. \cite{Christlieb2004,Frebel2018}).

In this work, the impact of the new atomic mass values has been studied for a binary neutron star merger scenario. The impact on the calculated $r$-process abundances is illustrated here by a representative dynamical ejecta trajectory for a 1.35 solar mass neutron-star merger from Ref.~\cite{mendozatemis2015}, with a very low initial electron fraction $Y_e=0.016$ and low entropy per baryon $s/k_B=8$.  Most of the prompt ejected mass (up to 90\%) is assumed to originate from these types of reheated, fission-recycling trajectories which all yield very similar abundances with the mass model used. Thus, the results are largely independent of the specific astrophysical conditions.

The $r$-process simulation followed the procedure outlined in \cite{Mumpower2016}. In the simulations, the entropy per baryon increased to $s/k_B \approx 100$ due to nuclear reheating. The initial timescale was around 40 ms, after which a homologous expansion was assumed \cite{mendozatemis2015}. The baseline calculations were done with the experimental mass values from AME16 \cite{AME16} supplemented by theoretical FRDM12 \cite{Moller2012} mass model values where no experimental data existed. The second calculation included the JYFLTRAP mass values from the first measurement campaign \cite{vilen2018}. Finally, the third set included all JYFLTRAP measurements performed in this region (\cite{vilen2018} and this work). To be consistent, calculated and experimental mass values were not combined in the calculation of a given $S_n$ value. The neutron-capture rates were calculated with the commonly used TALYS code \cite{talys} with the three different mass datasets described above. For fission product distributions, a simple asymmetric split \cite{Mumpower2017} was assumed. This ensured that the produced fission fragments fall into the $\mathrm{A}=130$ region and the rare-earth peak forms entirely via the dynamical formation mechanism of Refs.~\cite{Surman1997,mumpower2012_RE}. The branching ratios and $\beta$-decay half-lives were taken from NUBASE16 \cite{nubase16} or Ref.~\cite{moller2003}. 

\begin{figure}[tpb]
 	\begin{center}
     	\centerline{\includegraphics[trim = 5mm 0mm 5mm 5mm, clip,width=\linewidth]{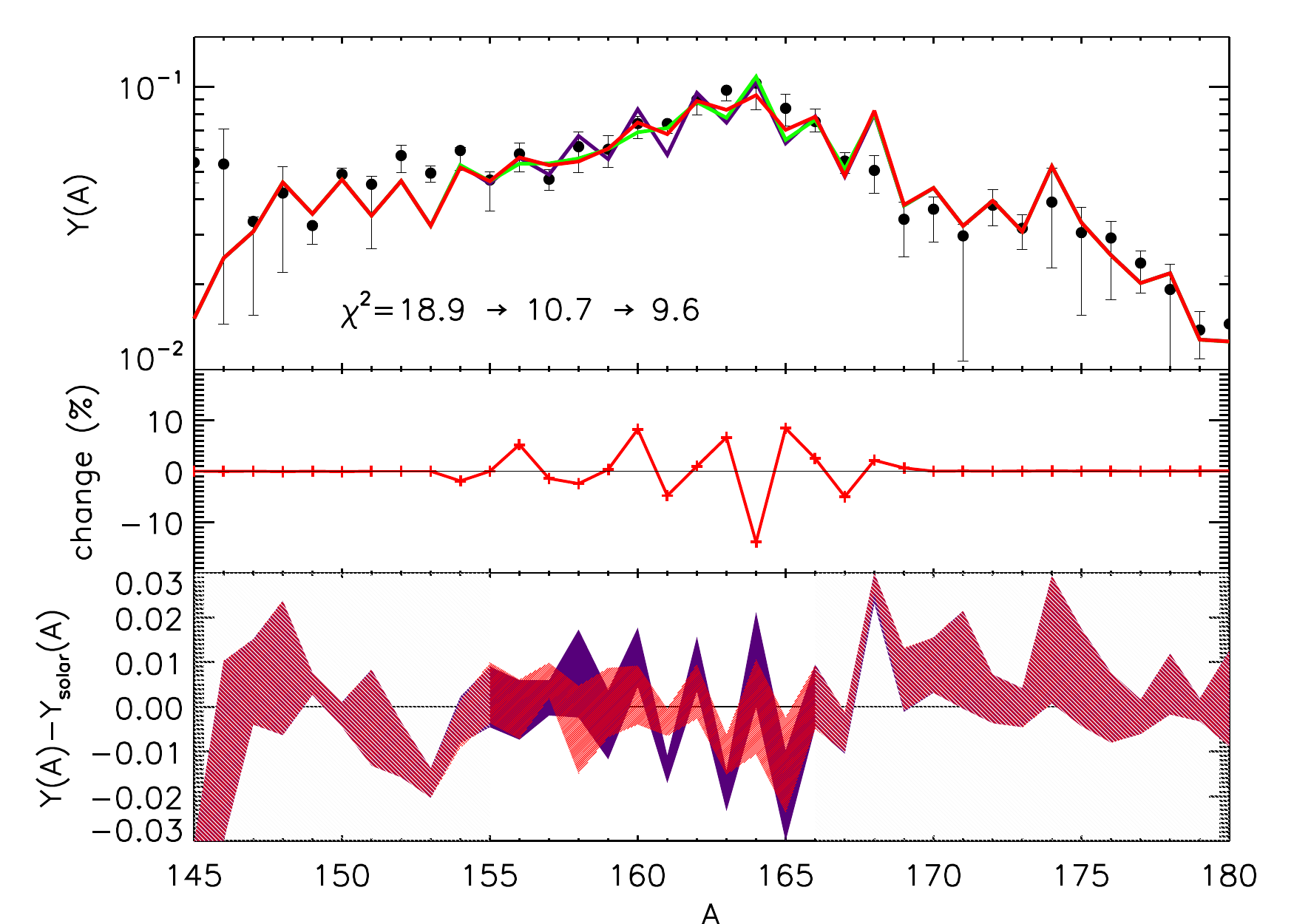}}
     	\caption{(Color on-line) Top: Solar $r$-process abundances $Y$ as a function of the mass number $A$ (black data points) together with the calculated abundances using the AME16 \cite{AME16} and FRDM12 \cite{Moller2012} mass-excess values (purple), and with the addition of the values from the first experimental campaign at JYFLTRAP (green) \cite{vilen2018}, and with the masses from this work (red). Middle: Change (in \%) between the calculated abundances from this work and from the previous JYFLTRAP experiment \cite{vilen2018}. Bottom: Residuals based on the mass values from this work (red) and the baseline (purple), where the bands represent the solar abundance uncertainties.}
     \label{fig:rprocessplot}
 	\end{center}
 \end{figure}

Figure~\ref{fig:rprocessplot} shows the impact of the masses determined in this work and \cite{vilen2018} on the calculated $r$-process abundances. The masses from \cite{vilen2018} were shown to severely affect the abundance pattern, resulting in a smoothening and better agreement with the solar abundances. The new masses from this work still affect the abundance pattern resulting in less staggering and a smoother profile at the top of the abundance peak. The reduced staggering at the summit of the peak might be caused by the reduced odd-even staggering in the $S_n$ values, now clearly seen in Fig.~\ref{fig:Dn} for the various chains measured. Such a reduced staggering reveals an increasing over-prediction by the FRDM mass model, especially for the Tb and Gd isotopic chains (see Fig.~\ref{fig:Dn-diff}). 
Furthermore, the small reduction in the following metric \(\chi^2 = \sum{ \{[Y(A)_{solar} - Y(A)_{calc}]/\sigma[Y(A)_{solar}] \}^2}\) from 10.7 to 9.6 indicates that the new masses further improved the matching with solar abundance data. The bottom panel of Fig.~\ref{fig:rprocessplot} shows the dramatic improvement in the matching between the solar abundance and the calculated abundance once the new JYFLTRAP masses from this work and \cite{vilen2018} are included. Finally, the middle panel shows that the new JYFLTRAP mass measurements change the calculated abundances up to 10$\%$ as compared to the results from \cite{vilen2018}.

\section{\label{sec:conclusions}Conclusions}

The masses of 13 nuclides in the rare-earth region have been measured using the JYFLTRAP Penning trap facility. This second campaign of measurements adds to the 12 masses previously measured in this sensitive region by JYFLTRAP \cite{vilen2018} and the 10 masses measured by the CPT \cite{hartley2018,orford2018}. In this second JYFLTRAP campaign, the masses of eight isotopes, namely $^{161}$Pm, $^{163}$Sm, $^{164,165}$Eu, $^{167}$Gd, and $^{165,167,168}$Tb were measured for the first time. The mass of $^{166}$Tb was found to be consistent with the AME16 value, while being 54 times more precise. The result for $^{154}$Nd agreed well with the recent PI-ICR measurement from CPT \cite{orford2018}. Due to previous disagreements and inconsistencies, the masses of $^{162,163}$Eu and $^{163}$Gd have been remeasured. We confirm both the ground-state mass and the isomeric-state energy of $^{162}$Eu measured by the CPT. The re-measurement of $^{163}$Gd and $^{163}$Eu using calibrant ions $^{136}$Xe and $^{133}$Cs, respectively, indicates that the reference ion used for these isotopes in the first JYFLTRAP campaign \cite{vilen2018} was incorrectly assigned, and was most likely $^{146}$La$^{16}$O$^{1}$H$^+$, or its mixture with $^{163}$Dy$^+$. Using the mass of this molecule as a calibrant leads to mass-excess values consistent with the CPT measurements for $^{163}$Gd and $^{163}$Eu. 

The impact of the new mass values on nuclear structure was studied via different parameters and derivatives of the mass surface. All mass models, including FRDM, were found to over-predict the odd-even staggering in $S_n$ values. The new mass values reveal an unusual enhancement in two-neutron shell-gap energies at $N$ = 97 for the odd-odd nuclei $^{162}$Tb. Such an enhancement is also seen at $N$ = 93 for $^{154}$Pm, but is absent at $N$ = 95 for $^{158}$Eu. Finally, a similar enhancement in the $\delta V_{pn}$ values is also seen for $^{154}$Pm and $^{162}$Tb, while being muted for $^{158}$Eu. Further studies are needed to understand the nature of the unusual enhancement in $D_{2n}$ value for the odd-odd nuclei $^{154}$Pm and $^{162}$Tb and for the sudden quenching in $\delta V_{pn}$ values for $^{158}$Eu.

The mass measurements presented in this work provide essential nuclear data for the $r$-process calculations, complementing the knowledge of the rare-earth region together with the recent beta-decay half-life measurements from the Radioactive Isotope Beam Factory \cite{wu2017}. While the new mass-excess values are shown to further reduce the staggering at the top of the rare-earth abundance peak in the $r$-process, the effect is more modest than previously seen \cite{vilen2018}. In the future, the isomeric states in the region can be further explored with the PI-ICR technique which has been demonstrated to work well for $^{162}$Eu in this work. More mass measurements in this region are warranted to refine theoretical mass models used for the $r$-process calculations as well as to better understand this region with rapid changes in nuclear structure. 

\acknowledgments{This work has been supported by the Academy of Finland under the Finnish Centre of Excellence Programme 2012-2017 (Nuclear and Accelerator Based Physics Research at JYFL) and by the National Science Foundation (NSF) Grants No. PHY-1713857. A.K., D.N., L.C., and T.E. acknowledge support from the Academy of Finland under projects No. 275389 and 295207. M.M. carried out this work under the auspices of the National Nuclear Security Administration of the U.S. Department of Energy at Los Alamos National Laboratory under Contract No. DE-AC52-06NA25396. R.S. work is funded in part by the DOE Office of Science under contract DE-SC0013039. The authors would like thank S. Frauendorf and I. Bentley for fruitful discussions.}

%

\end{document}